\documentclass[a4paper]{article} 

\usepackage{amssymb, amsthm, amsmath}
\usepackage{hyperref}
\usepackage{authblk}
\usepackage{verbatim}
\usepackage{enumerate}
\usepackage{url}
\usepackage{setspace} %doublespacing

\usepackage{graphicx} % COMPILES WITH FIGURES

\usepackage{physics} % for bra, ket

\usepackage[comma,authoryear]{natbib} 
\usepackage[margin=1in]{geometry}

\begin{document}
 
% =====================================
% DOUBLE SPACING
% =====================================
 %\doublespacing
% =====================================

 \begin{flushleft}
{\Large
\textbf{Quantum decision theory in simple risky choices}
}
\\
M. Favre$^{1,\ast}$, 
A. Wittwer$^{4}$,
H.R. Heinimann$^{5}$,
V.I. Yukalov$^{1,3}$,
D. Sornette$^{1,2}$ 
\\
\bf{1} ETH Z\"urich, Department of Management, Technology and Economics, Scheuchzerstrasse 7, 8092 Z\"urich, Switzerland
\\
\bf{2} Swiss Finance Institute, c/o University of Geneva, 40 blvd. Du Pont d'Arve, CH-1211 Geneva 4, Switzerland
\\
\bf{3} Bogolubov Laboratory of Theoretical Physics, Joint Institute for Nuclear Research, Dubna 141980, Russia
\\
\bf{4} Collegium Helveticum, University of Zurich and ETH Zurich, Schmelzbergstrasse 25, CH-8092 Zurich, Switzerland
\\
\bf{5} ETH Z\"urich, Future Resilient Systems at the Singapore-ETH Centre (SEC)
\\
$\ast$ Corresponding author: maroussiafavre@ethz.ch 
\end{flushleft}

%\tableofcontents

%\newpage

\abstract{
Quantum decision theory (QDT) is a recently developed theory of decision making based on the mathematics of Hilbert spaces, a framework known in physics for its application to quantum mechanics. This framework formalizes the concept of uncertainty and other effects that are particularly manifest in cognitive processes, which makes it well suited for the study of decision making. QDT describes a decision maker's choice as a stochastic event occurring with a probability that is the sum of an objective utility factor and a subjective attraction factor. QDT offers a prediction for the average effect of subjectivity on decision makers, the quarter law. We examine individual and aggregated (group) data, and find that the results are in good agreement with the quarter law at the level of groups. At the individual level, it appears that the quarter law could be refined in order to reflect individual characteristics. This article revisits the formalism of QDT along a concrete example and offers a practical guide to researchers who are interested in applying QDT to a dataset of binary lotteries in the domain of gains.
}

\vspace{0.3cm}

{\bf JEL classification:} C44, D03, D71, D81, D83

\vspace{0.3cm}

{\bf Keywords:} Decision making, quantum decision theory, risk, uncertainty, binary lotteries, gains, how-to guide

\section*{Introduction}

In this article, we apply quantum decision theory (QDT) to a dataset of choices between a certain and a risky lottery, both in the domain of gains. Each decision task consists in a choice between (1) a risky option in which the decision maker can win a fixed amount, here $y=50$ CHF with probability $p$, or nothing with probability $1-p$, and (2) a certain option in which the decision maker gets an amount x with certainty, where $0<x<y$. For example, a typical question could be to choose between the following options: ``option 1: get CHF $50$ with a $40$\% chance (or nothing with a $60$\% chance); option 2: get CHF $10$ for sure." Our dataset includes the $200$ decisions performed by each of $27$ subjects with various values of $p$ and $x$.

%We use this dataset to illustrate how to apply QDT. This theory of decision making offers a quantitative prediction, the so-called quarter law, which concerns the average effect of subjectivity on people's decisions. In the present case, the quarter law allows us to predict the frequency with which either of two options will be chosen. We examine individual and aggregated (group) results, and find that our results are in good agreement with the quarter law at the level of groups. At the individual level, it appears that the quarter law could be refined in order to reflect individual characteristics. We also perform a gender analysis suggesting that men in our sample are more risk-taking than women, on average. However, the small sample size does not allow to generalize this result to gender differences in general. We present these results in order to illustrate how QDT can be used to examine aggregates of decision makers.

QDT offers a quantitative prediction, the so-called quarter law, which concerns the average effect of subjectivity on people's decisions. In the present case, the quarter law allows us to predict the frequency with which either of two options will be chosen. 

With this article, our aim is to provide a practical how-to guide for QDT in its present state, not to resolve paradoxes of human decision making (as has been done in QDT articles by \citealp{Processing2009, Math2010, Decision2011}) or extend the theory. QDT publications \citep{Quantum2008, Physics2009, Processing2009, Scheme2009, QDTEntangl2010, Math2010, Decision2011, 2013Multimode, 2014Conditions, QDTManip14, 2013SelfOrg, QDTInfoSocial14} stress that the theory is general and applies to any decision problem. Accordingly, these articles abstractly derive the quarter law from a general framework. Here, we choose to consider a simple, concrete decision problem and remain within its frame to obtain the quarter law. We hope that this presentation of the theory helps the reader understand it in details. This is one contribution of this article.

Another contribution is to offer a practical illustration of how to apply QDT to a dataset of binary lotteries in the domain of gains, so as to test the quarter law at the level of individuals and groups when each subject is faced with a number of different decisions. Previous QDT publications, e.g. \citet{Processing2009, Decision2011}, typically test the quarter law on datasets in which individual results are bundled into a single statistics. Namely, in general, all that is known is the proportion of subjects who chose an option. By contrast, in this dataset, the choices of each subject in $200$ different decision tasks are available. Future experiments can thus use the methodology we develop here. Through this effort, we seek to address requests expressed by attendants of QDT conference talks. In this respect, \citet{Calibration2016} may also be consulted. 

We examine individual and aggregated (group) results, and find that our results are in good agreement with the quarter law at the level of groups. At the individual level, it appears that the quarter law could be refined in order to reflect individual characteristics. We use gender to distribute subjects into two groups and apply QDT to compare these groups and test the quarter law within each group. Gender is a convenient variable on which to test the aggregation of individual data. Indeed, it is a readily available individual characteristics that splits the sample in two sets of almost equal size. Moreover, there is a vast literature on gender-specific risk-taking behavior \citep{Eagly95,Byrnes99,EckelGrossman08,CrosonGneezy09}. The gender analysis we perform suggests that men in our sample are more risk-taking than women, on average, and that there is a higher inter-individual variability among women than men. However, the small sample size does not allow to generalize these results to gender differences in general. We present this analysis in order to illustrate and clarify how QDT can be used to examine aggregates of decision makers.

Let us stress that QDT does not suppose that the brain is a quantum object. The mathematical formalism of quantum mechanics, rigorously developed by \citep{Neumann55}, is that of complex separable Hilbert spaces. This formalism was already recognized by the founders of quantum mechanics to be well suited to describe the processes of human decision making \citep{Bohr29, Bohr33, Bohr37, Neumann55}. Indeed, it naturally integrates the notion of uncertainty. QDT represents cognitive states and prospects as vectors in a Hilbert space. The idea that a system is in a superposition of states until it gets measured parallels the way our cognitive state is indefinite until we take a decision. Moreover, vectors in a Hilbert space can be entangled, as can be the options among which we choose. We may identify feelings, subconscious processes or contextual effects with so-called hidden (inaccessible) variables. Finally, from the mathematics of Hilbert spaces naturally derive order effects and non-additive probabilities incorporating interference terms. 
%\footnote{We do not, as of yet, attempt to answer whether QDT can actually be reproduced by a theory of hidden variables.} 

We thus use the same mathematical formalism as that of quantum mechanics, but do not claim that neurological processes are quantum in nature. For clarity, let us give a parallel example. Newton (1642-1726/7) was interested in planetary motion and developed calculus to establish the laws of movement and gravitation. That does not mean that there is anything fundamentally planetary to calculus. This formalism very generally allows the study of change. In the case of planetary motion, calculus describes the change of position and speed along physical trajectories in space and time. Meanwhile, calculus has been used in every scientific domain concerned with change. In the same way, the mathematical framework of Hilbert spaces is known in physics for its application to quantum mechanics, but it can be said to generally apply to the study of uncertainty, and this is what motivates us to make use of it for decision making processes. 

Others have recently followed a similar path, most notably Busemeyer and collaborators (e.g. \citealp{Busemeyer06,BusemeyerTrueblood11,Busemeyer15}); see \citet{Ashtiani15} for a broad review of current so-called quantum-like models of cognition. QDT is constructed as a complete, general framework, can be applied to any problem in decision theory and offers quantitative predictions \citep{Ashtiani15}.

Let us now give a few elements of decision theory that help understand how QDT relates to classical theories of that field that are well known in the fields of psychology and behavioral economics. For this review, we consider how classical theories evaluate simple gambles. Simple gambles, as \citet{KahnemanBook} (p.269) puts it, are to decision theorists what fruit flies are to geneticists. A simple gamble is a lottery that yields, for example, a 20\% chance to win CHF 50 and a 80\% chance to gain nothing. Throughout this paper, we call such gambles ``prospects, ``options" or ``lotteries." A lottery (indexed by $j$) is characterized by a set of outcomes and their probability of occurrence, and is written as
\begin{equation}
L_j = \{ x_n, p_j(x_n) : n=1,2,...,N\},
\end{equation}
where $p_j$ is a probability measure over the set of payoffs $\{x_n\}$ and thus belongs to the interval $[0,1]$ and is normalized to one. 

In experimental settings, it is common for decision makers to have to choose between two simple gambles or between a gamble and a sure thing, as in ``would you rather have a 20\% chance to win CHF 50 (with an 80\% chance to get nothing), or get CHF 10 for sure?". We call this a ``decision task" (alternatively, ``choice problem", or ``game"). We write the former example as
\begin{eqnarray}
L_1 &=& \{ 50, 0.2 | 0, 0.8 | 10, 0\}, \label{ex1} \\
L_2 &=& \{ 50, 0 | 0, 0 | 10, 1\}, \label{ex2}
\end{eqnarray}
where the vertical line separates different outcomes, and all amounts present in the decision task appear in both lotteries.

In decision theory, such decision tasks serve as simple models for the more complex decisions we face in everyday life. In real situations, outcomes are often not expressed in monetary terms, and their probabilities are unknown. For instance, we are unable to assign a probability to the event that our happiness will increase if we get a new job and move, nor do we assign an exclusively quantitative value to this change of happiness. Nevertheless, it is the consensus of the field that simple choice problems such as the above elicit risk preferences, and we will focus on lotteries of that form. 

A few risk elicitation methods are reviewed by \citet{2008Harrison}. The method used to generate our dataset is called ``random lottery pair design" in the former review. This method, made popular by \citet{94HeyOrme}, consists in offering the decision maker a series of random lottery pairs in sequence. Our dataset was produced as part of a PhD thesis \citep{AmreiThesis} (chap. 7) and the tool created for that purpose was called by its authors Randomized Lottery Task (RALT).

In accordance with probability theory, \citet{Pascal1670} expressed the idea that it would be rational to choose the option with the highest expected value, given by the weighted sum
\begin{equation} \label{utility_pascal}
U(L_j) = \sum_n x_n p_j(x_n).
\end{equation}

Later on, \citet{Bernoulli1738} examined the relationship between the psychological and the objective value of money. He introduced the idea of a decreasing marginal value of wealth, which amounts to introducing a non-decreasing and concave utility function $u$ transforming values in (\ref{utility_pascal}), giving an expected utility:
\begin{equation} \label{utility_bernoulli}
\tilde{U}(L_j) = \sum_n u(x_n) p_j(x_n).
\end{equation}
Bernoulli proposed that $u(x)=\ln(x)$, such as to reflect that we become more indifferent to changes of wealth as the initial amount of wealth increases. For example, an increase of wealth from 1 to 4 million has a higher psychological value than an increase from 4 to 7 million. 

%ref-check von Neumann capital
The view that people should and will consistently choose the option with the highest expected utility given by Eq~(\ref{utility_bernoulli}) (with different forms of $u(x)$) remained dominant for a very long time. \citet{NeumannMorgenstern53} expressed axioms of rational behavior according to which rationality consists in maximizing an expected utility. Expected utility theory was the foundation of the rational-agent model, whereby people are represented by rational and selfish agents exhibiting tastes that do not change over time. This forms the foundation of the standard economic approach to decisions under risk and uncertainty \citep{Gollier04}. 

Pioneered by \citet{1979KT}, prospect theory modifies expected utility theory in order to explain a collection of observations showing that people exhibit a variety of cognitive biases contradicting rationality, in particular the Allais paradox \citep{Allais53}. In prospect theory, the utility $\hat{U}(L_j)$ is given by
\begin{equation} \label{utility_PT}
\hat{U}(L_j) = \sum_{n=1}^{N} v(x_n) w(p(x_n))~,
\end{equation}
where the value function $v(x)$ is constructed based on a reference point so that relative (and not absolute) wealth variations are considered in the expected utility $\hat{U}(L_j)$. \citet{1992TverskyKahn} proposed the following parametric functions (interpreted below): \begin{equation} \label{prospect_value}
v(x) = \left\{
    \begin{array}{rl}
      x^\alpha, &  x \ge 0; \\
      -\lambda (-x)^\beta, &  x < 0.
    \end{array} \right.
\end{equation}
 \begin{equation} \label{prospect_weight}
 w^{+}(p) = \frac{p^\gamma}{(p^\gamma + (1-p)^\gamma)^{1/\gamma}}, \quad  w^{-}(p) = \frac{p^\delta}{(p^\delta + (1-p)^\delta)^{1/\delta}}~.
 \end{equation}
Note that the two latter expressions distinguish between gains ($+$) and losses ($-$).

Other functional specifications have been formulated in so-called non-expected utility theories (for a review, see \citealp{2006Stott}). The underlying ideas behind the value and the probability weighting functions is that people exhibit diminishing sensitivity to the evaluation of changes of wealth in the domain of gains (encapsulated in the concavity of $v(x)$ for $x \ge 0$), the reverse effect in the domain of losses (loss aversion), and a subjective probability weighting that overweights small probabilities and underweights large probabilities (and can be different for gains and losses). Also, evaluation is relative to a neutral reference point (here $x=0$), below which changes of wealth are seen as losses. This contrasts with standard utility theory, in which only the final state of wealth contributes to the utility evaluation. Parameters (here $\alpha, \beta, \gamma, \delta$) are fitted to experimental data to reflect the choices of groups or individuals, assuming that people prefer the option with the highest utility. 

\citet{MostellerNogee51} observed that decision making can be inherently stochastic, that is, the same decision maker may take a different decision when faced with the same question at different times. \citet{Luce59} formalized this idea by offering a form for the probability of choosing one option over another. Several functions have since been proposed for this so-called stochastic specification, stochastic error, or choice function; for reviews see \citet{2006Stott} or \citet{2008Harrison}. A probabilistic element can thus be integrated in prospect theory models, providing in principle a limit for the explanation and prediction power of such models (e.g. \citealp{94HeyOrme, Harless94} and \citealp{Murphy14}). 

Prospect theory is a fertile field of research that currently dominates decision theory \citep{WakkerBook}. However, some scholars \citep{SafraSegal08,NajjarWeinstein09} point out that non-expected utility theories in general do not remove paradoxes, create inconsistencies and always necessitate an ambiguous fitting that cannot always be done. In this context, QDT proposes an alternative perspective that may provide novel insights into decision making.

The main features of QDT are the following. First, as mentioned above, QDT derives from a complete, coherent theoretical framework that explicitly formalizes the concepts of uncertainty, entanglement and order effects. In the framework of QDT, paradoxes of classical decision theory such as the disjunction effect, the conjunction fallacy, the Allais paradox, the Ellsberg paradox or the planning paradox (to name just a few) find quantitative explanations \citep{Processing2009, Math2010, Decision2011}. Within QDT, behavioral biases result from interference caused by the deliberations of decision makers making up their mind \citep{QDTEntangl2010}.

Second, QDT is inherently a probabilistic theory, in the sense that it focuses on the fraction of people who choose a given prospect, or on the frequency with which a single decision maker does so over a number of repetitions of the same question. In the theoretical part of this article, we derive the following expression for the probability $p(L_j)$ that a prospect $L_j$ be chosen by one or several decision makers:
\begin{equation} \label{proba_QDT}
p(L_j) = \frac{U(L_j)}{\sum_i U(L_i)} + q(L_j),
\end{equation}
where $U(L_j)$ is the same utility as in (\ref{utility_pascal}) and $q(L_j)$ is the attraction factor; in this term are encompassed the ``hidden variables" of decision theory, i.e. feelings, contextual factors, subconscious processes, and so on. Eq~(\ref{proba_QDT}) can be derived from seeing prospects and cognitive states as vectors in a Hilbert space, as we show in the next section.

Last, QDT distinguishes itself from classical and other quantum-like decision theories by offering quantitative predictions. Specifically, the quarter law predicts that the average absolute value of $q(L_j)$ is $1/4$ under the null hypothesis of no prior information.

Besides explaining several paradoxes on binary lotteries \citep{Processing2009, Math2010, Decision2011}, QDT has been developed in several directions. It provides expressions for discount functions, employed in the theory of time discounting \citep{TimeDiscount02} and explains dynamical inconsistencies \citep{Physics2009}. QDT also describes the influence of information and a surrounding society on individual decision makers \citep{QDTInfoSocial14,QDTManip14}. While QDT has been developed to describe the behavior of human decision makers, it can also be used as a guide to create artificial quantum intelligence \citep{Scheme2009}. 

This article is structured as follows. The next section presents the theoretical formulation of QDT in the formalism of Hilbert spaces. In particular, we recall the derivation of the quarter law, which gives the average amplitude of the attraction factors. In the methods, we formulate how QDT applies to the dataset under study. We then expose our main results comparing QDT with experiments. Finally, we discuss our results and conclude.

\section*{Model: quantum decision theory} \label{section-qdt}

\subsection*{Hilbert space formalism} \label{qdt_comp_prosp}

This section shows how Eq~(\ref{proba_QDT}) can be derived from seeing prospects and a decision maker's state of mind as vectors in a complex Hilbert space in the simple, concrete case of a choice between two lotteries in the domain of gains. This section requires some knowledge of the mathematics of Hilbert spaces and quantum mechanics. After Eq~(\ref{proba_QDT}) is derived, the rest of this article does not necessitate any such knowledge.

\subsubsection*{Prospects}

We consider the ``fruit fly" of decision theory \citep{KahnemanBook} (p. 269), i.e. the situation where subjects can choose between two lotteries in the domain of gains, denoted by $L_1$ and $L_2$ and expressed as
\begin{eqnarray}
L_1 &=& \{ x_n, p_1(x_n) : n=1,2,...,N\}, \\ 
L_2 &=& \{ x_n, p_2(x_n) : n=1,2,...,N\}.
\end{eqnarray}
An example is given by Eqs~(\ref{ex1}) and (\ref{ex2}).

\subsubsection*{Observables and Hilbert spaces}

Let us introduce two observables A and B, represented by operators $\hat{A}$ and $\hat{B}$ acting on Hilbert spaces $\mathcal{H}_A$ and $\mathcal{H}_B$, respectively. Each observable can take on two values,
\begin{equation}
A = \{ A_1, A_2 \} \quad \textrm{and} \quad B = \{ B_1, B_2 \}.
\end{equation}
$A_1$ and $A_2$ represent the two options presented to the subjects of the experiment: when the lottery $L_j$ is chosen, $A$ takes the corresponding value $A_j$ ($j=1,2$). $B$ embodies an uncertainty \citep{QDTManip14} (p. 1162). For instance, $B_1$ (resp., $B_2$) can represent the confidence (resp., the disbelief) of the decision maker that she correctly understands the empirical setup, that she is taking appropriate decisions, and that the lottery is going to be played out as announced by the experimenter. $B$ thus encapsulates the idea that these choices involve an uncertainty for the decision maker, although it is not necessarily explicitly expressed in the setup. The eigenstates of each operator form an orthonormal basis of the associated Hilbert space,
\begin{equation}
\mathcal{H}_A = \textrm{span} \{ \ket{A_1}, \ket{A_2} \}, \quad \mathcal{H}_B = \textrm{span} \{ \ket{B_1}, \ket{B_2} \}.
\end{equation}
A decision maker's state is defined in the tensor-product space
\begin{equation}
\mathcal{H}_{AB} = \mathcal{H}_A \otimes \mathcal{H}_B,
\end{equation}
spanned by the tensor-product states $\ket{A_i} \otimes \ket{B_j}$ ($i,j \in \{1,2\}$). For simplicity, we write $\ket{A_i B_j}$ without the tensor product sign:
\begin{equation}
\mathcal{H}_{AB} = \textrm{span} \{ \ket{A_1 B_1}, \ket{A_1 B_2}, \ket{A_2 B_1}, \ket{A_2 B_1} \}.
\end{equation}

\subsubsection*{States}

A decision maker is assumed to be in a \emph{decision-maker state} $\ket{\psi} \in \mathcal{H_{AB}}$, which is a linear combination of the basis states:
\begin{equation} \label{psi_coeff}
\ket{\psi} = \alpha_{11} \ket{A_1 B_1} + \alpha_{12} \ket{A_1 B_2} + \alpha_{21} \ket{A_2 B_1}  + \alpha_{22} \ket{A_2 B_2},
\end{equation}
with $\alpha_{mn} \equiv \alpha_{mn}(t) \in \mathbb{C}$ for $m,n \in \{1,2\}$. Each decision maker is characterized by his or her own set of time-dependent, non-zero coefficients $\{\alpha_{mn}(t)\}$, $m,n \in \{1,2\}$. This superposition state reflects our indecision until we make a choice. The time dependence implies that the same individual may take different decisions when asked the same question at different times. The time evolution can be seen as due to endogenous processes in the decision maker's body and mind (e.g. breathing, digestion, feelings, thoughts) as well as exogenous factors (interactions with the surroundings). 

We assume that $\ket{\psi}$ is normalized, i.e.
\begin{equation} \label{psi_norm}
\langle \psi | \psi \rangle = |\alpha_{11}|^2 + |\alpha_{12}|^2 + |\alpha_{21}|^2 + |\alpha_{22}|^2 = 1
\end{equation}

The \emph{prospect states} are defined as product states $\ket{\pi_j} \in \mathcal{H_{AB}}$,
\begin{eqnarray}  \label{pi_j_def}
\ket{\pi_j} &=& \ket{A_j} \otimes \left\{ \gamma^{j}_{j1} \ket{B_1} + \gamma^{j}_{j2} \ket{B_2} \right\} \\
&=& \gamma^{j}_{j1} \ket{A_j B_1} + \gamma^{j}_{j2} \ket{A_j B_2}, \quad \gamma^{j}_{kl} = 0, \quad \forall j,k \neq j,l \in \{1,2 \},
\end{eqnarray}
where $\gamma_{kl}^{j} \in \mathbb{C}$, $j,k,l \in \{1,2\}$. Concretely,
\begin{eqnarray}
\ket{\pi_1} &=& \gamma^{1}_{11} \ket{A_1 B_1} + \gamma^{1}_{12} \ket{A_1 B_2}, \quad \gamma^1_{21} = 0, \quad \gamma^1_{22} = 0 \\
\ket{\pi_2} &=& \gamma^{2}_{21} \ket{A_2 B_1} + \gamma^{2}_{22} \ket{A_2 B_2}, \quad \gamma^2_{11} = 0, \quad \gamma^2_{12} = 0.
\end{eqnarray}
The prospect states are superposition states, each corresponding to the decision maker ultimately choosing either lottery with indefinite mixed feelings about the setup and context. The use of three indices in $ \gamma^{j}_{kl}$ is useful to distinguish options that are available to the decision maker as opposed to options that are excluded, as illustrated in the next section.

\subsubsection*{Probability measure \label{nthgw}}

QDT assumes that the following process takes place: the decision maker approaches any question in a state $\ket{\psi}$ of the form of Eq~(\ref{psi_coeff}). During her deliberations, she transits to either $\ket{\pi_1}$ or $\ket{\pi_2}$. If she transits to $\ket{\pi_j}$ ($j=1,2$), she chooses the lottery $L_j$. The experimenter observes the choice of the lottery $L_j$, but does not know whether $B$ took the value $B_1$ or $B_2$. The decision maker herself may remain undecided about $B_1$ or $B_2$. After the decision, the decision-maker state becomes $\ket{\psi'}$, a superposition state with the form of (\ref{psi_coeff}), but possibly with different coefficients than before, due to the time evolution. The same process is repeated when the next question is asked.

Hence, we identify the probability that the lottery $L_j$ be chosen with the probability that $\ket{\psi}$ makes a transition to the superposition state $\ket{\pi_j}$ ($j=1,2$):
\begin{equation} \label{p_transition}
p(L_j) \approx \textrm{proba}(\psi \rightarrow \pi_j) = | \langle \psi | \pi_j \rangle |^2 \quad (j=1,2) ~.
\end{equation}
The approximation sign is there because, a priori, one does not have $| \langle \psi | \pi_1 \rangle |^2 + | \langle \psi | \pi_2 \rangle |^2  = 1$. This is because, formally, there are additional states in $\mathcal{H}_{AB}$ than just $\ket{\pi_1}$ and $\ket{\pi_2}$ to which the system can make a transition; for instance, $\ket{\pi_3} = \gamma^{3}_{11} \ket{A_1 B_1} + \gamma^{3}_{12} \ket{A_1 B_2} + \gamma^{3}_{21} \ket{A_2 B_1} $. However, $\ket{\pi_3}$ does not correspond to any decision that can be made in the experimental setting. In practice, decision makers have to choose between $L_1$ and $L_2$ (alternatively, if they fail to answer properly, their data is discarded). This amounts to constraining the system such that
\begin{equation} \label{p_sumto1}
\sum_{j=1}^2 p(L_j) = 1,
\end{equation}
which can be achieved by defining $p(L_j)$ as follows:
\begin{equation} \label{p_def_norm}
p(L_j) := \frac{| \langle \psi | \pi_j \rangle |^2}{| \langle \psi | \pi_1 \rangle |^2 + | \langle \psi | \pi_2 \rangle |^2}. 
\end{equation}
By defining the normalization quantity $P$, the former expression can also be written as
\begin{equation} \label{P_def}
p(L_j) = \frac{1}{P} | \langle \psi | \pi_j \rangle |^2; \quad P = | \langle \psi | \pi_1 \rangle |^2 + | \langle \psi | \pi_2 \rangle |^2.
\end{equation}
Eq~(\ref{p_def_norm}) ensures that $\sum_{j=1}^2 p(L_j) = 1$ and that $p(L_j) \in [0,1]$ ($\forall j=1,2$).

The probability $p(L_j)$ ($j=1,2$) is to be understood as the theoretical frequency of the prospect $L_j$ being chosen over a large number of times. Two types of setups can be put in place. The first setup involves a number of agents; in that case $p(L_j)$ gives the fraction of agents expected to choose $L_j$. In the second setup, a single decision maker is faced with a number of decision tasks, among which the same decision task appears several times. This setup is designed such that each repetition of the same decision task can be assumed to be independent of previous occurrences. That is, we assume negligible memory and influence of previous decisions. In this second setup, the option $L_j$ is chosen a fraction $p(L_j)$ of the time by the same decision maker. QDT treats similarly these two setups.

\subsubsection*{Utility and attraction factors} \label{S-utility-attr-fact}

Let us examine the quantity $| \langle \psi | \pi_j \rangle |^2$ in terms of the coefficients given in (\ref{psi_coeff}) and (\ref{pi_j_def}):
\begin{eqnarray} 
| \langle \psi | \pi_j \rangle |^2 &=& \langle \psi | \pi_j \rangle \langle \pi_j | \psi \rangle 
= (\alpha^{*}_{j1} \gamma^{j}_{j1} + \alpha^{*}_{j2} \gamma^{j}_{j2}) (\alpha_{j1} \gamma^{j*}_{j1} + \alpha_{j2} \gamma^{j*}_{j2}) \\
&=& |\alpha_{j1}|^2  |\gamma^{j}_{j1}|^2 + |\alpha_{j2}|^2  |\gamma^{j}_{j2}|^2 + \alpha^{*}_{j1} \gamma^{j}_{j1}  \alpha_{j2} \gamma^{j*}_{j2} + \alpha^{*}_{j2} \gamma^{j}_{j2} \alpha_{j1} \gamma^{j*}_{j1}. \label{proba_coeff}
\end{eqnarray}
In quantum mechanics, the last two terms in (\ref{proba_coeff}) correspond to an interference between different intermediate states as the system makes a transition from $\ket{\psi}$ to $\ket{\pi_j}$. In decision theory, this interference can be understood as originating from the decision maker's deliberations as she is in the process of weighting up options and making up her mind. These interference terms are encapsulated into what is called in QDT the \emph{attraction factor},
\begin{equation} \label{qcoeff}
q(L_j) = \frac{1}{P} (\alpha^{*}_{j1} \gamma^{j}_{j1}  \alpha_{j2} \gamma^{j*}_{j2} + \alpha^{*}_{j2} \gamma^{j}_{j2} \alpha_{j1} \gamma^{j*}_{j1}),
\end{equation}
where $P$ is the normalization quantity introduced in (\ref{P_def}). The attraction factor is interpreted in QDT as a contextual object encompassing subjectivity, feelings, emotions, cognitive biases and framing effects. 

The classical terms become the so-called \emph{utility factor},
\begin{equation} \label{fcoeff}
f(L_j) = \frac{1}{P} (|\alpha_{j1}|^2  |\gamma^{j}_{j1}|^2 + |\alpha_{j2}|^2  |\gamma^{j}_{j2}|^2).
\end{equation}
Eq~(\ref{p_def_norm}) can then be rewritten as
\begin{equation} \label{pfq}
p(L_j) = f(L_j) + q(L_j).
\end{equation}

In the absence of interference effects, only the utility factor $f(L_j)$ remains in $p(L_j)$. When $q(L_j)=0$, the standard classical correspondence principle \citep{1920Bohr} leads one to let $p(L_j)$ connect to more classical decision theories such as those mentioned in the introduction. The utility factor $f(L_j)$ has to take the form of a probability function, satisfying
\begin{equation} \label{f_conditions}
\sum_{j} f(L_j) = 1 \quad \textrm{and} \quad  f(L_j) \in [0,1] . 
\end{equation}

As the simplest non-parametric formulation connecting $f(L_j)$ to classical decision theories, the utility factor is equated with
\begin{equation}
f(L_{j}) = \frac{ U(L_{j}) } { \sum_i U(L_{i})}, \label{fdef}
\end{equation}
where $U(L_{j})$ is an expected value or expected utility, as in (\ref{utility_pascal}), (\ref{utility_bernoulli}) or (\ref{utility_PT}). In this article, we use the simple non-parametric form given by (\ref{utility_pascal}), $U(L_{j})=\sum_{n=1}^{N} p(x_{n}) x_{n}$, making it an expected value. 

A more general form for $f(L_{j})$ has been proposed \citep{2013SelfOrg}, namely
\begin{equation}
f(L_{j}) = \frac{ U(L_{j}) \exp \{ \beta U(L_{j}) \} } { \sum_i U(L_{i}) \exp \{ \beta U(L_{i}) \} },\label{fbeta}
\end{equation}
where $\beta \geq 0$ is called the ``belief parameter." Previous applications of QDT to empirical data (e.g. \citealp{Math2010, Decision2011}) use the case $\beta=0$, which is also what we use in this article. 

Let us examine the implications of conditions (\ref{f_conditions}) when the utility factors are written as in (\ref{fcoeff}). Imposing that the utility factors sum to one means that 
\begin{equation} \label{f_sumstoP}
|\alpha_{11}|^2  |\gamma^{1}_{11}|^2 + |\alpha_{12}|^2  |\gamma^{1}_{12}|^2 + |\alpha_{21}|^2  |\gamma^{2}_{21}|^2 + |\alpha_{22}|^2  |\gamma^{2}_{22}|^2 = P.
\end{equation}
Based on its definition in (\ref{P_def}) and (\ref{proba_coeff}), the normalization coefficient $P$ is explicited as
\begin{eqnarray}
P &=& |\alpha_{11}|^2  |\gamma^{1}_{11}|^2 + |\alpha_{12}|^2  |\gamma^{1}_{12}|^2 + |\alpha_{21}|^2  |\gamma^{2}_{21}|^2 + |\alpha_{22}|^2  |\gamma^{2}_{22}|^2  + \\
&+& \alpha^{*}_{11} \gamma^{1}_{11}  \alpha_{j2} \gamma^{1*}_{12} + \alpha^{*}_{12} \gamma^{1}_{12} \alpha_{11} \gamma^{1*}_{j1}
+ \alpha^{*}_{21} \gamma^{2}_{21}  \alpha_{22} \gamma^{2*}_{22} + \alpha^{*}_{22} \gamma^{2}_{22} \alpha_{21} \gamma^{2*}_{21}.
\end{eqnarray}
From the definition of the attraction factors (\ref{qcoeff}), and using (\ref{f_sumstoP}), this can be rewritten as
\begin{equation}
P = P + P(q(L_1) + q(L_2)),
\end{equation}
which yields
\begin{equation} \label{altern}
q(L_1) + q(L_2) = 0,
\end{equation}
since the attraction factors are not null by definition. The constraint that the utility factors sum to one thus imposes that the attraction factors sum to zero. The latter condition is the focus of the next section.

\subsection*{Quarter law: prediction of QDT on attraction factors} \label{section_quarterlaw}

In this section, we recall the derivation of the so-called quarter law, which governs the typical amplitude of the attraction factors (e.g. \citealp{QDTManip14}, p.1159). 

Since $p(L_j)$ and $f(L_j)$ both sum to one over the $L=2$ prospects, it derives from (\ref{pfq}) that the attraction factors $q(L_j)$ sum to zero. Also, since $p(L_j)$ and $f(L_j)$ are both comprised in the interval $[0,1]$, $q(L_j)$ lies in the interval $[-1,1]$. These two conditions are called in QDT the 
\emph{alternation conditions}:
\begin{equation} \label{alternation}
-1 \leq q(L_j) \leq 1, \quad \sum_{j=1}^L q(L_j) = 0.
\end{equation}

Arising from interference effects between prospects in the mind of the decision maker, the attraction factor can be seen as a random quantity varying across decision makers and over time for a single decision maker. Thus, the choices made by different individuals or by the same person at different times constitute a number of realizations of $q$ equal to $L$ (the number of prospects in each decision task) multiplied by the number of choices made. Let us introduce $\varphi(q)$, the normalized distribution of $q$, and write
\begin{equation} \label{varphi}
\int_{-1}^{1} \varphi(q) dq = 1.
\end{equation}

Because of the alternation conditions (\ref{alternation}), in the presence of $L=2$ prospects, the distribution of $q$ is symmetrical around zero and the mean of $q$ is zero:
\begin{equation}
\int_{-1}^{1} \varphi(q) q dq = 0.
\label{ghethnw}
\end{equation}

Let us define the quantities
\begin{equation} \label{qplusminus}
q_+ = \int_{0}^{1} \varphi(q) q dq, \qquad q_-= \int_{-1}^{0} \varphi(q) q dq.
\end{equation}

From the alternation conditions (\ref{alternation}) leading to (\ref{ghethnw}), we have
\begin{equation}
q_+ + q_- = 0.
\end{equation}

In the absence of any information, the variable $q$ is equiprobable, i.e. the distribution $\varphi(q)$ is uniform. From (\ref{varphi}), it derives that $\varphi(q)=\frac{1}{2}$. Put into (\ref{qplusminus}), this yields
\begin{equation} 
q_+ = \frac{1}{4}, \quad q_- = -\frac{1}{4}.
\end{equation}

Applied to a binary lottery choice, this yields the prediction that, when $L_{1}$ is the most attractive prospect, we have
\begin{equation} \label{quarter}
 \bar{q}(L_{1}) = \frac{1}{4}, \quad \bar{q}(L_{2}) = -\frac{1}{4},
\end{equation}
where the signs are reversed if $L_{2}$ is the most attractive prospect. In QDT, this is called the \emph{quarter law} and constitutes a quantitative prediction on the average value of the attraction factors in any given experiment, under the assumption of no additional prior information.

Several families of distributions $\varphi(q)$ yield the same average value of $1/4$ for the attraction factor, which makes this result quite general, as pointed out by \citet{QDTManip14}. For example, it is the case for the symmetric beta distribution,
\begin{equation}
\varphi(q) = \frac{\Gamma(2 \alpha)}{2 \Gamma^2(\alpha)} |q|^{\alpha-1} (1-|q|)^{\alpha-1},
\end{equation}
defined on the interval $q \in [-1,1]$, with $\alpha>0$ and $\Gamma(\alpha)$ the gamma function. The beta distribution is employed in many applications, for example in Bayesian inference as a prior probability distribution. The quarter law also follows from several other distributions normalized on the interval $[-1,1]$, such as the symmetric quadratic distribution,
\begin{equation}
\varphi(q) = 6 \left(|q| - \frac{1}{2}\right)^2,
\end{equation}
and the symmetric triangular distribution,
\begin{equation}
    \varphi(q) =
    \begin{cases}
      2|q| & \text{if } 0 \leq |q| \leq \frac{1}{2},\\
      2(1-|q|) & \text{if } \frac{1}{2} < q \leq 1.
    \end{cases}
\end{equation}

Typical values of the attraction factors are derived by \citet{Quantitative2016} in the general case of a decision involving more than two prospects.

\subsection*{Sign of attraction factor in the case of close utility factors} \label{closeutilYS13}

A rule has been expressed by \citet{QDTManip14} (p.1162), giving the sign of the attraction factor for a binary decision task with close utility factors.

This rule considers two lotteries
 \begin{eqnarray}
  L_{1} &=& \{ x_i, p_1(x_i) : i = 1,2, ... \}, \\
  L_{2} &=& \{ y_j, p_2(y_j) : j = 1,2, ... \}. 
 \end{eqnarray}
Note that not all of the same amounts appear in the two prospects.
 
The maximal and minimal gains are denoted as
\begin{eqnarray}
x_{max} = \sup_{i} \{ x_{i} \}, &\quad& x_{min} = \inf_{i} \{ x_{i} \}, \\
y_{max} = \sup_{j} \{ y_{j} \}, &\quad& y_{min} = \inf_{j} \{ y_{j} \}.
\end{eqnarray}

The gain factor $g(L_{1})$ of the first prospect, the risk factor $r(L_{2})$ of choosing the second prospect, and the quantity $\alpha(L_{1})$ are then defined as follows:
\begin{equation}
g(L_{1}) = \frac{x_{max}}{y_{max}}
\end{equation}
\begin{equation}
r(L_{2}) = \left\{
    \begin{array}{rl}
       \frac{p_{2}(y_{min})}{p_{1}(x_{min})}, & p_{2}(y_{min}) < 1 \\
       0, &   p_{2}(y_{min}) = 1
    \end{array} \right.
\end{equation}
\begin{equation}
\alpha(L_{1}) = g(L_{1}) r(L_{2}) - 1.
\end{equation}

Using these quantities, the sign of the first prospect attraction factor, $q(L_{1})$, is defined by the rule
\begin{equation} \label{e-sign-q}
\textrm{sgn q} (L_{1})  = \left\{
    \begin{array}{rl}
      +1, &  \alpha(L_{1}) > 0 \\
      -1, &  \alpha(L_{1}) \le 0 .
    \end{array} \right.
\end{equation}

\section*{Methods} \label{S-method}

\subsection*{Prospects}

In the dataset we are analyzing, subjects have to choose between one of two lotteries. One lotteries, $L_1$, is risky: subjects can win either CHF 50, with a probability p, or nothing, with a probability 1-p. In the other lottery $L_2$, subjects get the sure amount x in CHF (i.e., with probability 1). This can be seen as a ``certain lottery." The condition $0<x<50$ is always ensured, otherwise there would be no point in choosing the risky lottery. The two lotteries are written
 \begin{eqnarray}
L_1 &=&  \{ 0, 1-p | x, 0 | y, p \} , \label {RALT_pi1} \\
L_2 &=& \{ 0,0 | x, 1 | y, 0 \} .  \label {RALT_pi2}
 \end{eqnarray}
 with $0<x<y$ and in the experiment, $y=50$ CHF.
 
Note that the same amounts appear in both prospects. Writing $L_1$ as (\ref{RALT_pi1}) means that one can get $0$ with probability $1-p$,  $x$ with probability $0$ and $y$ with probability $p$. Similarly, writing $L_2$ as (\ref{RALT_pi2}) corresponds to gaining $0$ with probability $0$, $x$ with probability $1$ and $y$ with probability $0$. This way of expressing the two prospects captures the fact that decision makers compare the two lotteries as they are making up their mind.

\subsection*{Decision tasks and participants} \label{setup}

In this dataset, a decision task is characterized by the values of the two parameters $p$ (the probability of winning the fixed amount CHF $50$ in the risky lottery $L_1$) and $x$ (the guaranteed payoff in the certain lottery $L_2$). In the experiment that generated the dataset, pairs of $(p,x)$ values were randomly generated and the corresponding decision tasks ($L_1$, $L_2$) were submitted to subjects. 

Each of $27$ participants was given $200$ randomly generated decision tasks, in $4$ runs of $50$ questions each. This amounts to $5,400$ data points. Overall, the probability $p$ was varied between $0.05$ and $0.9$ by steps of $0.05$, and the sure amount $x$ between $0$ and $49$ by steps of $1$. At most $900$ different $(p,x)$ pairs could thus be generated.

The detailed method of the experiment in given by \citet{AmreiThesis} (chap. 7), including the description of conditions between which we do not differentiate in the present analysis.

The mean age of participants was $25.8 \pm 3.6$ years. Thirteen women (mean age $25 \pm 3.3$ years) and fourteen men (mean age $27 \pm 3.7$ years) were included in the study. The decisions were contextualized, in the sense that participants were asked to imagine a bank investment. Each decision was equally important, as participants were promised to receive, at the end of the experiment, the amount resulting from one randomly chosen decision task. The Ethics Committee of the Canton of Zurich approved the study. All participants provided written, informed consent and were compensated for their participation.

\subsection*{Application of QDT to the dataset} \label{qdt}

We now present how QDT specifically applies to the empirical dataset under study. The theoretical correspondence was carried out in the section that follows the introduction. Indeed, this section considers a decision problem whose structure corresponds to that of the decision tasks generating the dataset.
 
\subsubsection*{Utility factors} 

As stated earlier, we equate prospect utilities with expected values given by (\ref{utility_pascal}), hence
\begin{eqnarray} 
 U(L_{1}) &=& (1-p) 0 + 0 \cdot x + p \cdot 50 = 50p \label{utility1},  \\
 U(L_{2}) &=& 0 \cdot 0 + 1 \cdot x + 0 \cdot 50 = x \label{utility2}.
\end{eqnarray}

Then, according to (\ref{fdef}), the utility factors are given by
\begin{eqnarray}
 f(L_{1}) &=& \frac{ U(L_{1}) } { U(L_{1}) + U(L_{2})} = \frac{50p}{50p+x} \label{frisk}, \\
 f(L_{2}) &=& \frac{ U(L_{2}) } { U(L_{1}) + U(L_{2})} = \frac{x}{50p+x} \label{frisk2}.
\end{eqnarray}

\subsubsection*{Aggregation of decision tasks} \label{S-aggreg}

Since QDT is a probabilistic theory, a decision task must be offered several times in order to obtain an empirical choice frequency approximating $p(L_j)$. As is the case of other datasets available to researchers, the present dataset was not generated with QDT in mind. We suggest in the discussion how to design decision tasks more suitable for testing QDT. There are several ways to aggregate the decision tasks, of which the following:

\begin{itemize}
\item[(i)] Only aggregate decisions presenting the same $(p,x)$ values, i.e. the same utilities $U(L_1)$, $U(L_2)$. 
\item[(ii)]  Aggregate decisions with close $(p,x)$ values. For example, $(p=0.1, x=42)$ and $(p=0.1, x=45)$ present the same value of p and close values of x, and could be considered so similar as to represent the same decision task. 
\item[(iii)]  Aggregate decisions presenting the same utility factors $f(L_1)$, $f(L_2)$. For instance, the $(p,x)$ pairs $A(p=0.1, x=42)$, $B(p=0.1, x=45)$ and $C(p=0.05, x=22)$ give respectively $f(L^A_{1}) = 0.106$, $f(L^B_1) = 0.1$ and $f(L^C_1)=0.102$. Rounding these utility factors to the lowest $0.01$, the three games A, B and C are aggregated as three realizations of the same decision task characterized by a utility factor $f(L_{1})=0.1$.
\end{itemize}

The issue with only aggregating identical $(p,x)$ pairs (option (i)) is that the available empirical dataset does not always present a sufficient number of realizations of each $(p,x)$ pair to perform a meaningful probabilistic analysis. For instance, in our dataset, $10.7$\% of all $(p,x)$ pairs were offered only once; $11.5$\% were offered twice; $30.4$\% were offered three times or less. At most, one $(p,x)$ pair was offered $126$ times. Ignoring decision tasks with less than a minimum number of realizations is possible, but means giving up a substantial amount of data.

\begin{figure}[ht!]
\begin{center}
\includegraphics[width=.6\textwidth]{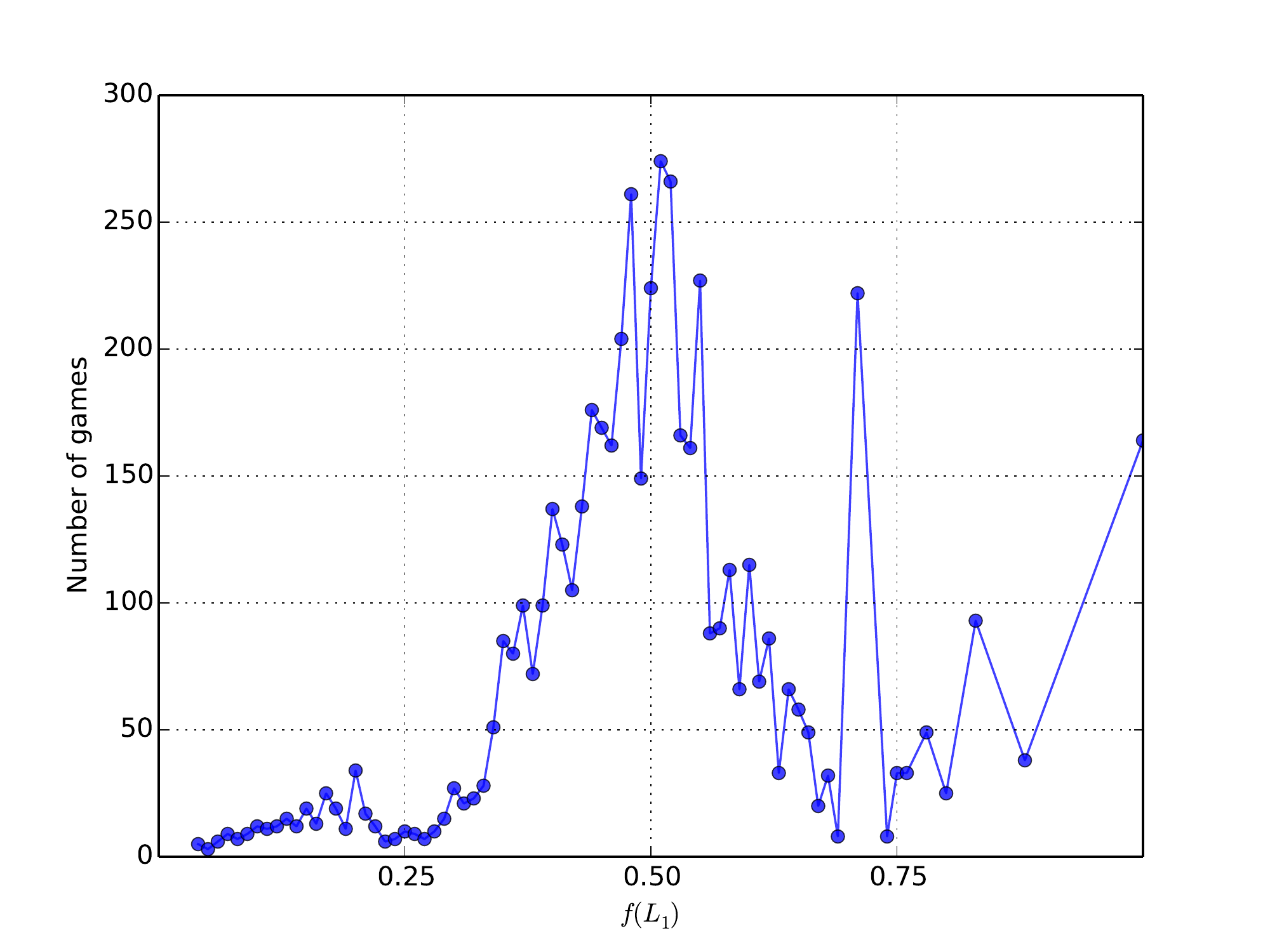}
\end{center}
\caption{Number of games offered in the dataset for each utility factor $f(L_1)$ rounded to the lowest 0.01.}
\label{Nb-games}
\end{figure}

Aggregating by utility factor (option (iii)) partly resolves this issue. When rounding $f(L_1)$ to the lowest $0.01$, no decision task characterized by its utility factor $f(L_1)$ was offered less than three times, and only $1.3$\% of all values of $f(L_1)$ were submitted exactly three times. At most, a given value of $f(L_1)$ was submitted $274$ times. The number of decision tasks given per value of $f(L_1)$ is shown in Fig~\ref{Nb-games}. It indicates that some decision tasks were oversampled while others were undersampled. In the discussion, we describe a sampling method that would be more adequate for our purposes.

%\begin{figure}[!h]
%\caption{{\bf Distribution of decision tasks.}
%Number of games offered in the dataset for each utility factor $f(L_1)$ rounded to the lowest $0.01$.}
%\label{Nb-games}
%\end{figure}

In option (ii), one has to decide how close values of $p$ and $x$ from different $(p,x)$ pairs have to be in order to aggregate the pairs. This option can bring an intermediate result between options (i) and (iii). In this analysis, we use option (iii).

\subsubsection*{Probability}

Empirically, we have access to the number of times $N_j$ a prospect was chosen by one or several decision makers over the $N$ times the same decision task was offered. Since the probabilities, utility factors and attraction factors for the two prospects ($j=1,2$) are related by Eqs~(\ref{p_sumto1}), (\ref{f_conditions}) and (\ref{altern}), the entire analysis can rest on the quantities associated with only one prospect. In what follows, we focus on the risky prospect, i.e. $j=1$.

The empirical frequency is given by
\begin{equation}
p_{\textrm{exp}}(L_1) = \frac{N_1}{N},
\end{equation}
which approximates the probability $p(L_1)$. In what follows, we identify $p_{\textrm{exp}}(L_1)$ with $p(L_1)$.

Let us introduce a random variable $X$ such that, at the $k$-th realization of a given decision task,
\begin{equation} \label{X_def}
X_k  = \left\{
    \begin{array}{rl}
      1, &  \textrm{if the risky prospect is chosen,} \\
      0, &  \textrm{if the certain prospect is chosen.} \\
    \end{array} \right.
\end{equation}
The probability $p(L_1)$ is thus the Bernoulli distribution of $X$. 
 
Let us additionally introduce the \emph{retract} function, which retracts a variable $z$ into an interval $[a,b]$:
\begin{equation}
\textrm{Ret}_{[a,b]}\{ z \}  = \left\{
    \begin{array}{rl}
      a, &  z \le a \\
      z, &  a < z < b \\
      b, & z \ge b
    \end{array} \right.
\end{equation}

Empirically, based on (\ref{pfq}), the following relationship holds between the probability $p(L_1)$, the utility factor $f(L_1)$ and the attraction factor $q(L_1)$: 
\begin{equation} \label{p_retract}
p(L_1)=\textrm{Ret}_{[0,1]} \left\{ f(L_1) + q(L_1) \right\},
\end{equation}
where the retract function ensures that $p(L_1)$ lies in $[0,1]$.

\subsubsection*{Attraction factors}

The attraction factor can be deduced from experimental results as
\begin{equation} \label{q_experim}
 q(L_1) = p(L_1) - f(L_1), 
\end{equation}
but only for $0<p(L_1)<1$, because of the retract function in (\ref{p_retract}). Indeed, for $p(L_1)=0$ and $p(L_1)=1$, $q(L_1)$ can have taken any value such that $f(L_1)+q(L_1) \leq 0$ or $f(L_1)+q(L_1) \geq 1$, respectively. Hence, the exact value of $q(L_1)$ that guided the decision maker(s) cannot be retrieved in these two cases; applying Eq~(\ref{q_experim}) may lead to over- or underestimate $q(L_1)$.

\begin{figure}[ht!]
\begin{center}
  \includegraphics[width=0.6\textwidth]{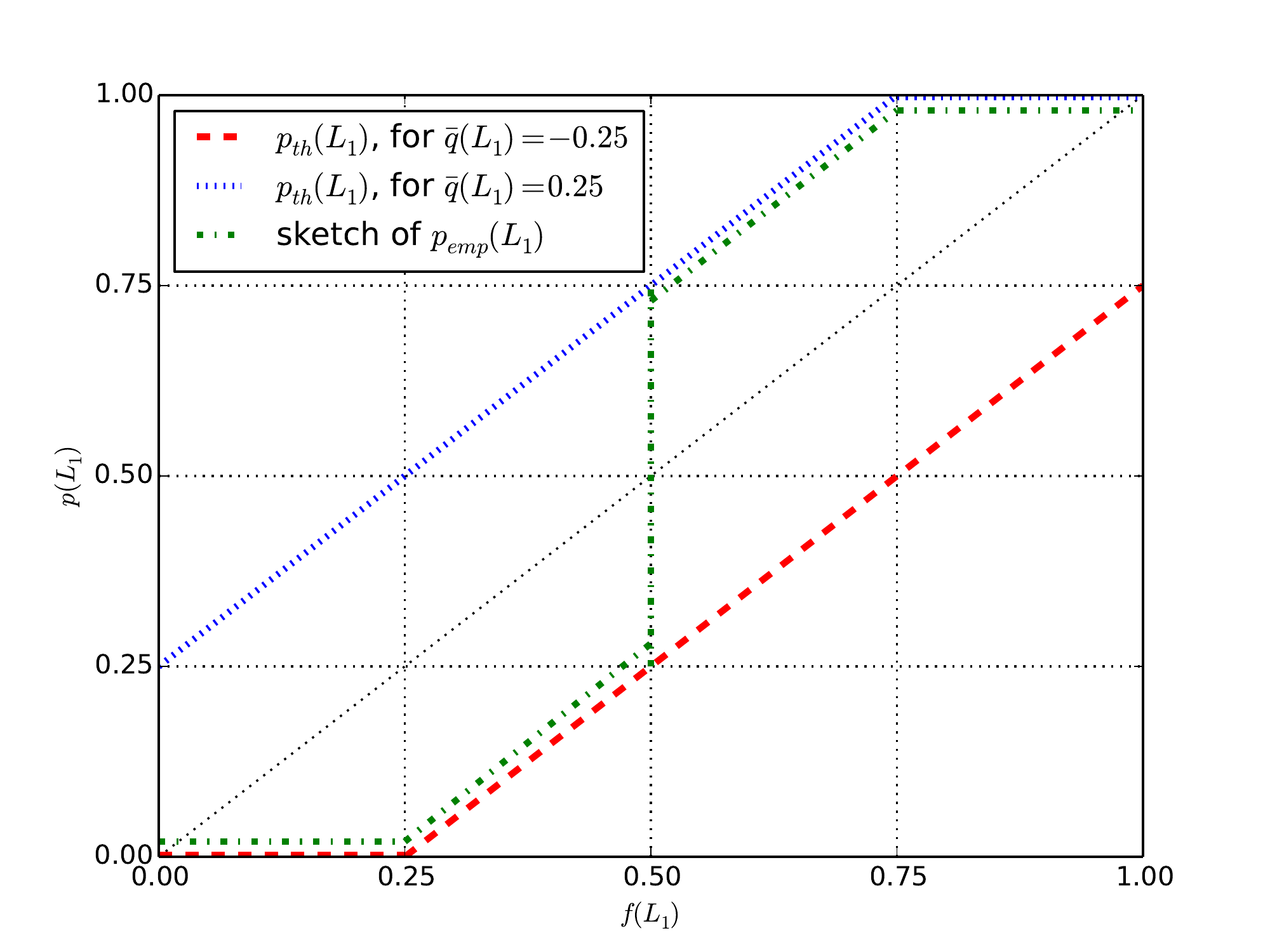}
\end{center}
\caption{Predicted (theoretical) appearance of the choice probability of the risky prospect, $p_{th}(L_1)$, for a positive (blue, dotted line) or negative (red, dashed line) attraction factor $\bar{q}(L_1)$ of absolute value equal, on average, to 0.25, and sketch of empirical probability, $p_{emp}(L_1)$ (green, dot-dash line).}
\label{QDT_prediction}
\end{figure}

The dataset consists in decision tasks with various values of $f(L_1)$ in $[0,1]$. In order to visualize to which extent empirical data conforms to the quarter law through Eq~(\ref{q_experim}), we plot empirical results as in Fig~\ref{QDT_prediction}, which shows $p(L_1)$ as a function of $f(L_1)$. Then $q(L_1)$ is given by the difference between $p(L_1)$ and the lower left to upper right diagonal of the graph ($p(L_1)=f(L_1)$), as long as $0<p(L_1)<1$. The absolute value of this distance is predicted by the quarter law to be equal to 0.25, on average. Table \ref{QDT_predict_table} and Fig~\ref{QDT_prediction} illustrate this prediction. Fig~\ref{QDT_prediction} also offers a sketch of our empirical results presented later.

%\begin{figure}[!h]
%\caption{{\bf QDT prediction (quarter law).}
%Predicted (theoretical) appearance of the choice probability of the risky prospect, $p_{th}(L_1)$, for a positive (blue, dotted line) or negative (red, dashed line) attraction factor $\bar{q}(L_1)$ of absolute value equal, on average, to 0.25, and sketch of empirical probability, $p_{emp}(L_1)$ (green, dot-dash line).}
%\label{QDT_prediction}
%\end{figure}

\begin{table}[!ht] 
\begin{center}
\begin{tabular}{|c|c|c|}
\hline
Region of $f(L_{1})$ & If $q(L_{1})=-0.25$ & If $q(L_{1})=0.25$ \\
\hline
 $0 \le f(L_{1}) < 0.25$ & $p(L_{1})=0$ &  $p(L_1)= f(L_{1}) + 0.25$ \\
 \hline
 $0.25 \le f(L_{1}) < 0.75$ & $p(L_1)= f(L_{1}) - 0.25$ &  $p(L_1)= f(L_{1}) + 0.25$ \\
 \hline
  $0.75 \le f(L_{1}) \le 0.1$ & $p(L_1)= f(L_{1}) - 0.25$ & 1 \\
  \hline
  \end{tabular}
  \end{center}
  \caption{Prediction on $p(L_1)$ by quarter law over different intervals of $f(L_1)$.}
   \label{QDT_predict_table}
 \end{table}

\subsubsection*{Sign of attraction factors}

The rule for the sign of the attraction factor in the case of close utility factors given by Eq~(\ref{e-sign-q}) applies to the lotteries written as
 \begin{eqnarray}
  L_{1} &=& \{ 50, p | 0, 1-p \}, \\
  L_{2} &=& \{ y, 1 \}.
 \end{eqnarray}
One gets $r(L_{2})=0$, $\alpha(L_{1})=-1$ and $\textrm{sgn q} (L_{1})=-1$. In other words, at and around $f(L_1) = f(L_2) = 0.5$, the risky prospect's attraction factor should be negative and the certain prospect should be preferred, in accordance with the principle of risk aversion. 

\subsubsection*{Summary}

For clarity, we summarize here the few steps necessary to obtain the attraction factors and test the quarter law in a dataset of choices between two lotteries in the domain of gains.
\begin{enumerate}
\item Choose a form for the utility factors satisfying $\sum_{j=1}^{2} f(L_j) = 1$ and $f(L_j) \in [0,1]$ $\forall j=1,2$.
\item Decide on an aggregation of decision tasks (e.g. per rounded value of utility factor).
\item For each aggregate, compute the frequency of a lottery $L_j$ being chosen, $p(L_j)$.
\item For $p(L_j)$ in $(0,1)$, compute for each aggregate the attraction factor $q(L_j)=p(L_j)-f(L_j)$.
\item Represent $p(L_j)$ as a function of $f(L_j)$.
\item Compare the absolute value of the attraction factor (at specific values of $f(L_j)$, or averaged over intervals of $f(L_j)$), $|q(L_j)|$, to the value $1/4$ predicted by the quarter law. 
\end{enumerate}

\section*{Results} \label{S-results}

\subsection*{Individual results}

\begin{figure}[ht!]
\begin{center}
 \includegraphics[width=.4\textwidth]{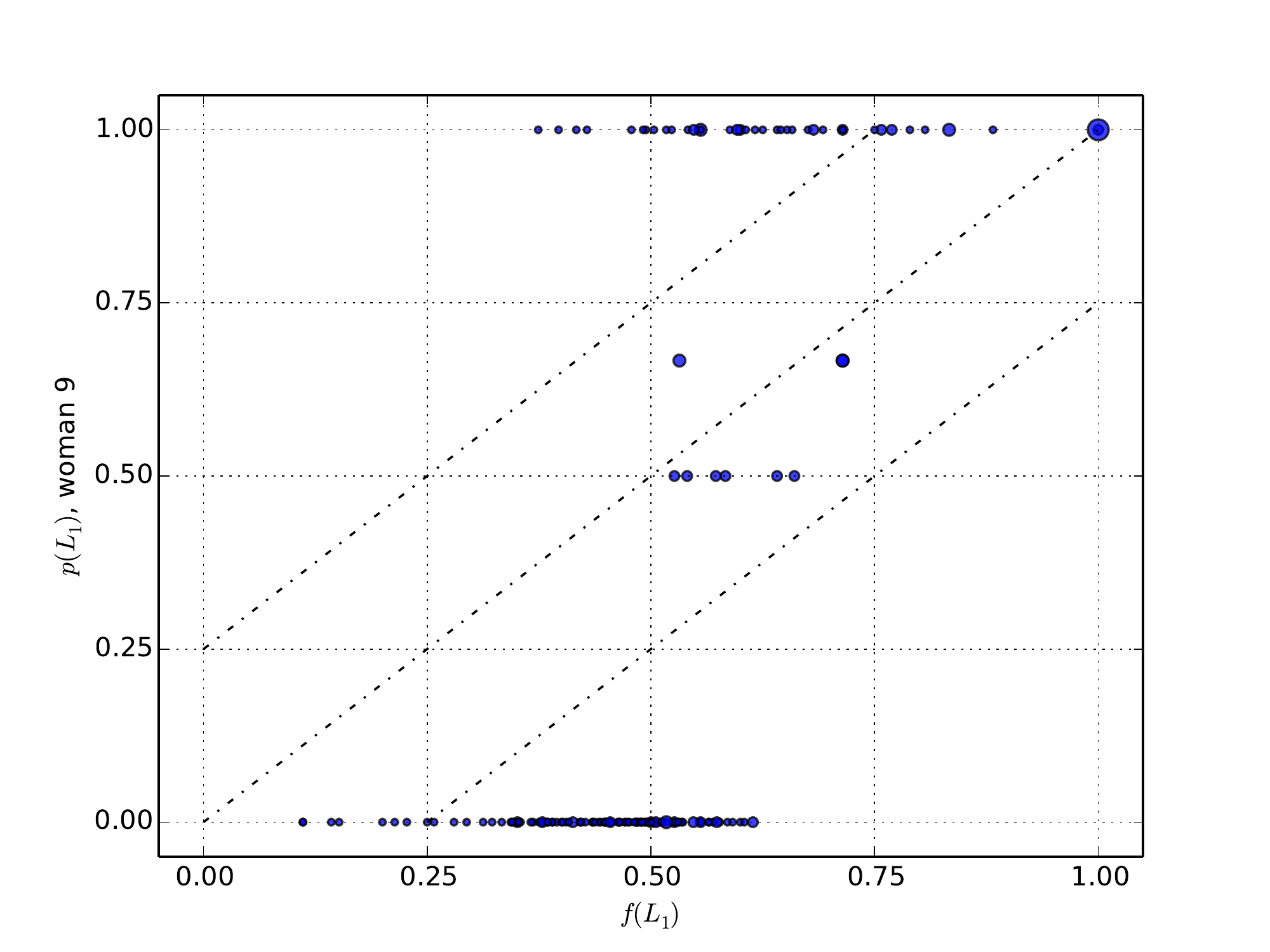} 
 \includegraphics[width=.4\textwidth]{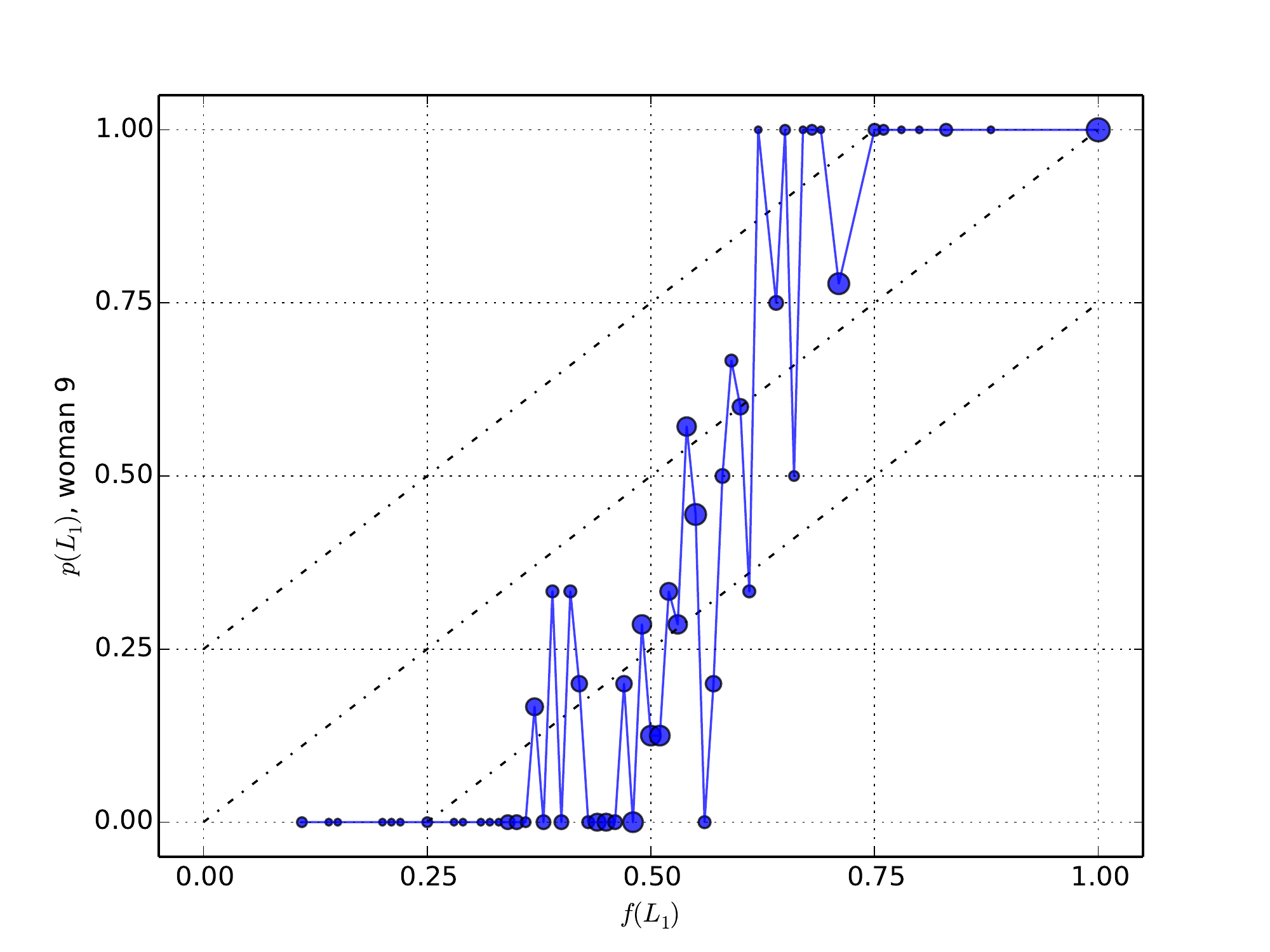}
\end{center}
\caption{For one individual, empirical probability $p(L_1)$ (choice frequency of the risky prospect) as a function of the utility factor $f(L_1)$. The area of the markers is proportional to the number of decision tasks per point (with a different proportionality factor in each panel). In the left panel, each point aggregates decision tasks that are exactly the same, i.e. presenting the same $(p,x)$ values and thus the same utilities $U_1$, $U_2$. The number of decision tasks per point varies from 1 to 4. In the right panel, each point aggregates decision tasks presenting the same value of $f(L_1)$ rounded to the lowest 0.01. There are between 1 and 16 decision tasks per point.}
\label{F-individuals-p-x}
\end{figure}

Fig~\ref{F-individuals-p-x} represents the results of one subject chosen for illustration purposes. We represent the frequency with which the individual chose the risky prospect, $p(L_1)$, as a function of the utility factor of this prospect, $f(L_1)$. In the left panel, each point comprises the results of decision tasks presenting the same $(p,x)$ values. The area of the markers is proportional to the number of decision tasks per point, which varies between $1$ and $4$. It is apparent in this panel that the majority of decision tasks were given only once, based on markers area as well as the fact that most points are situated at $p(L_1)=0$ or $p(L_1)=1$. 

% Place figure captions after the first paragraph in which they are cited.
%\begin{figure}[!h]
%\caption{{\bf Example of individual results for two different aggregations of decisions.}
%For one individual, empirical probability $p(L_1)$ (choice frequency of the risky prospect) as a function of the utility factor $f(L_1)$. The area of the markers is proportional to the number of decision tasks per point (with a different proportionality factor in each panel). In the left panel, each point aggregates decision tasks that are exactly the same, i.e. presenting the same $(p,x)$ values and thus the same utilities $U_1$, $U_2$. The number of decision tasks per point varies from $1$ to $4$. In the right panel, each point aggregates decision tasks presenting the same value of $f(L_1)$ rounded to the lowest $0.01$. There are between $1$ and $16$ decision tasks per point.}
%\label{F-individuals-p-x}
%\end{figure}

The left panel of Fig~\ref{F-individuals-p-x} illustrates the fact that, faced with decision tasks presenting the same utility factor $f(L_1)$, the same individual may not consistently choose the same prospect, especially at intermediary values of $f(L_1)$. Yet, at low and high values of $f(L_1)$, decisions tend to be consistent, as may be expected.

We observe a preference switch from the certain to the risky prospect as the utility factor of the risky prospect increases. This switch is well known in the behavioral economics literature. Its quantitative treatment usually consists in representing the frequency of a prospect being chosen (i.e. $p(L_1)$) as a function of a quantity depending on the question's parameters, which in our case is $f(L_1)$. To these data points is then typically fitted a logistic (S-shaped) function whose parameter can be interpreted as partly characterizing a decision maker's risk behavior (e.g. \citealp{MostellerNogee51, Murphy14}). The switch can be more or less sharp and is interpreted as decision makers being more or less discriminating and consistent. 

The right panel of Fig~\ref{F-individuals-p-x} also shows $p(L_1)$ as a function of $f(L_1)$ for the same individual, but here we aggregated $(p,x)$ values yielding the same $f(L_1)$ value rounded to the lowest $0.01$. As noted in the methods, there are thus more decision tasks per point, which motivates us to continue the analysis based on this aggregation. In contrast to aforementioned fitting approaches, we are interested in examining whether these data points follow the QDT prediction given by the quarter law, expressed by Eq~(\ref{quarter}). As a recall, the quarter law predicts that $p(L_1)$ is situated at $f(L_1) \pm 0.25$ on average.

\begin{figure}[ht!]
\begin{center}
 \includegraphics[width=.4\textwidth]{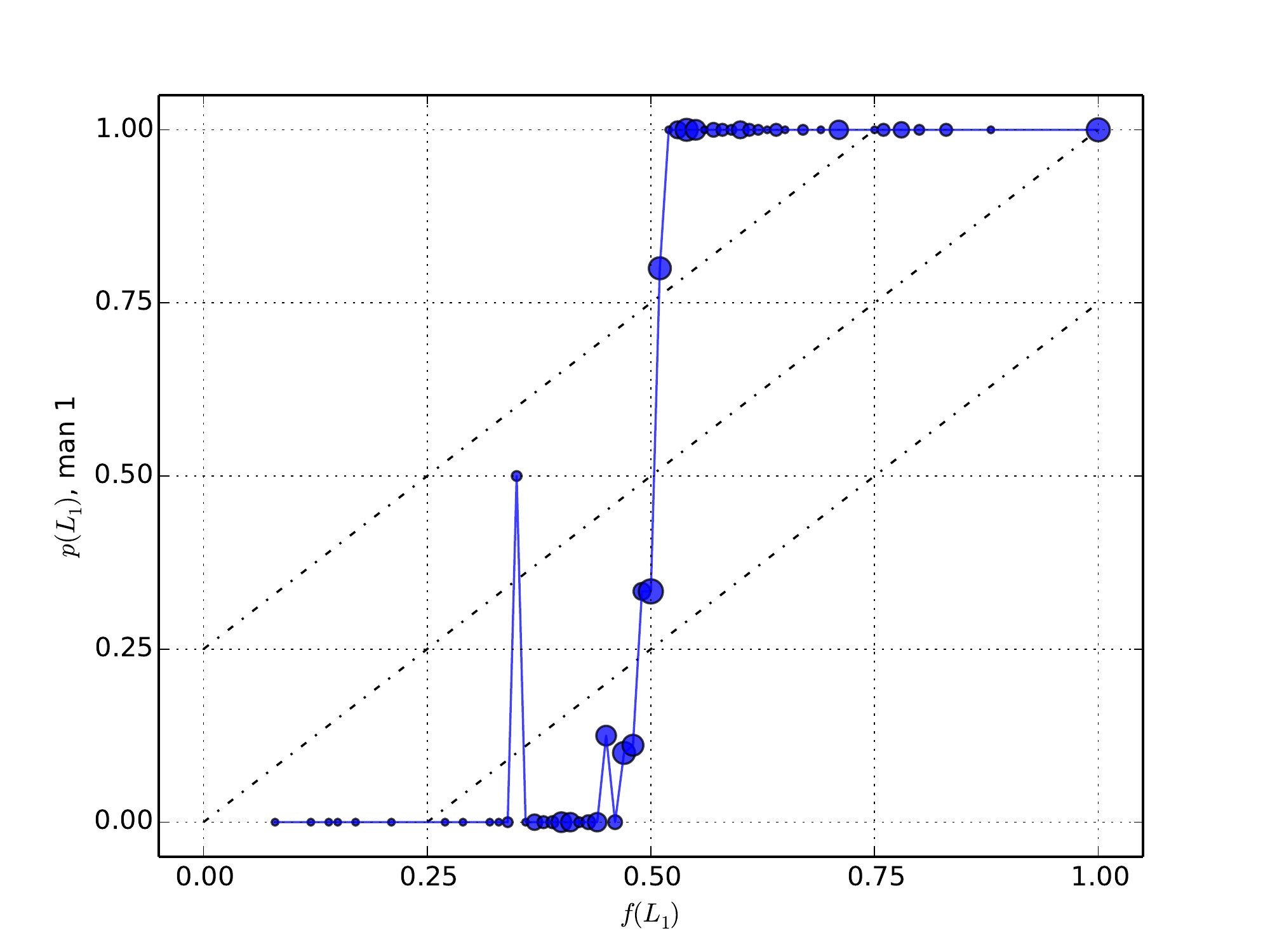} 
  \includegraphics[width=.4\textwidth]{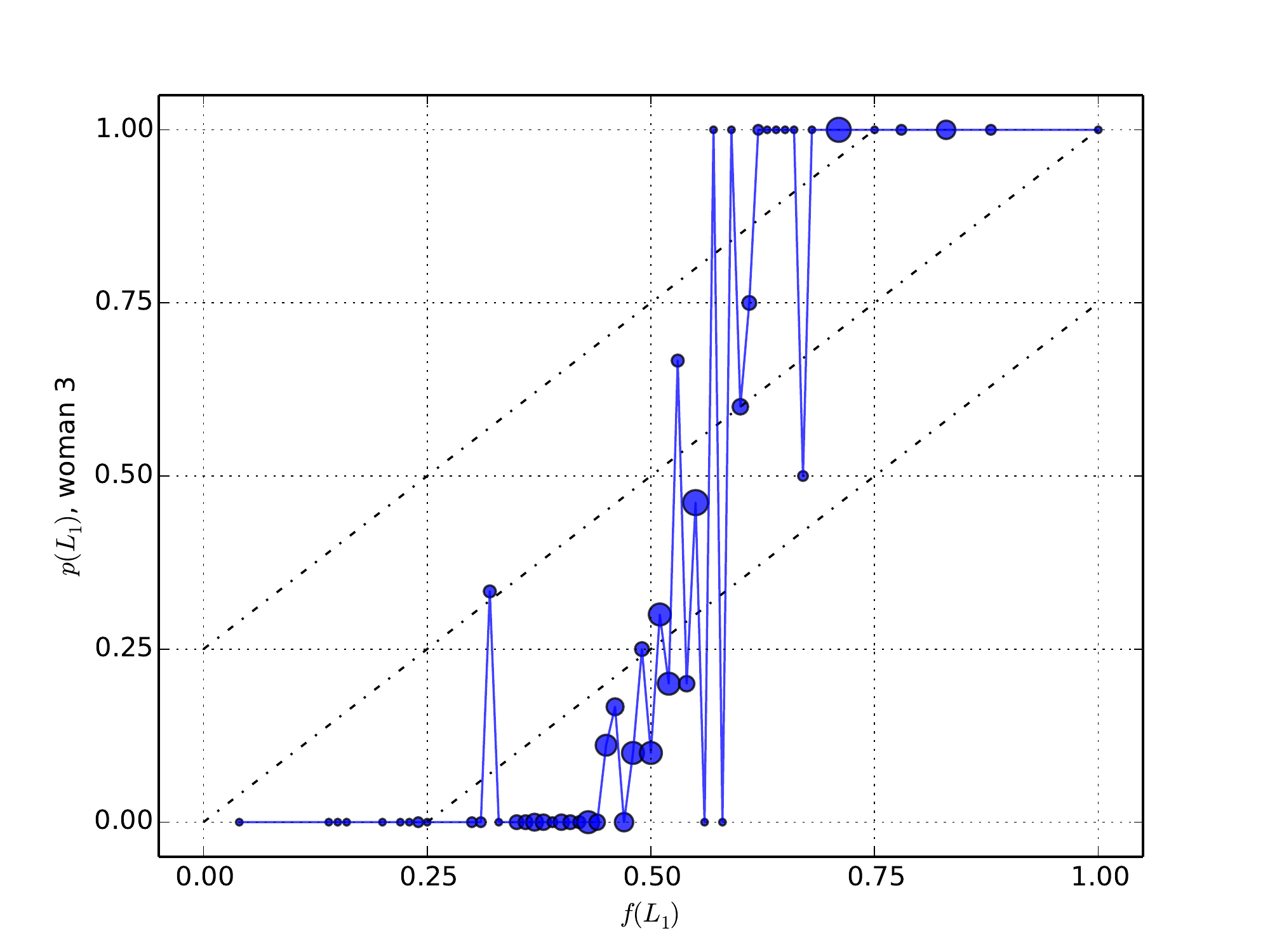} \\
    \includegraphics[width=.4\textwidth]{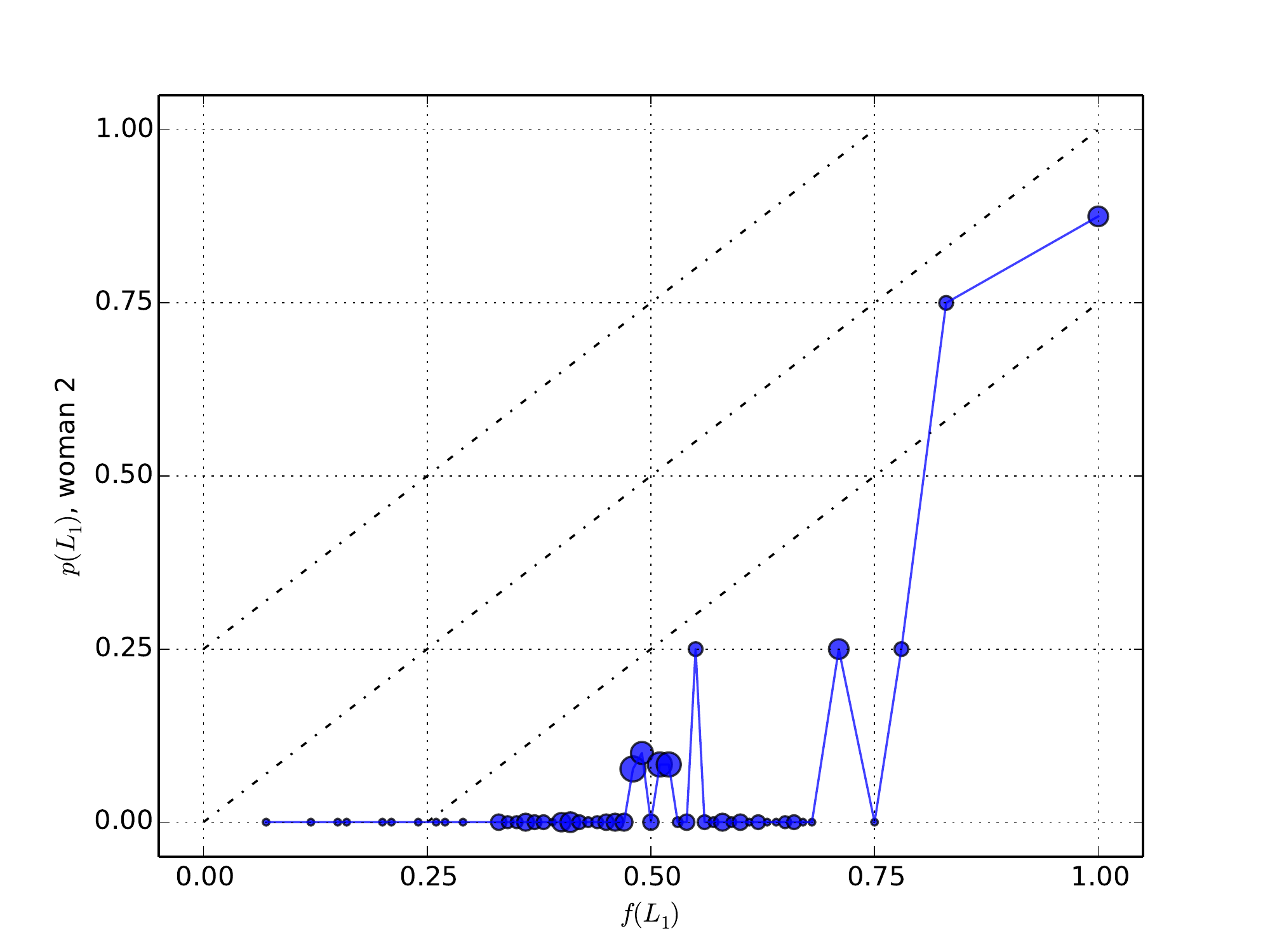}
 \includegraphics[width=.4\textwidth]{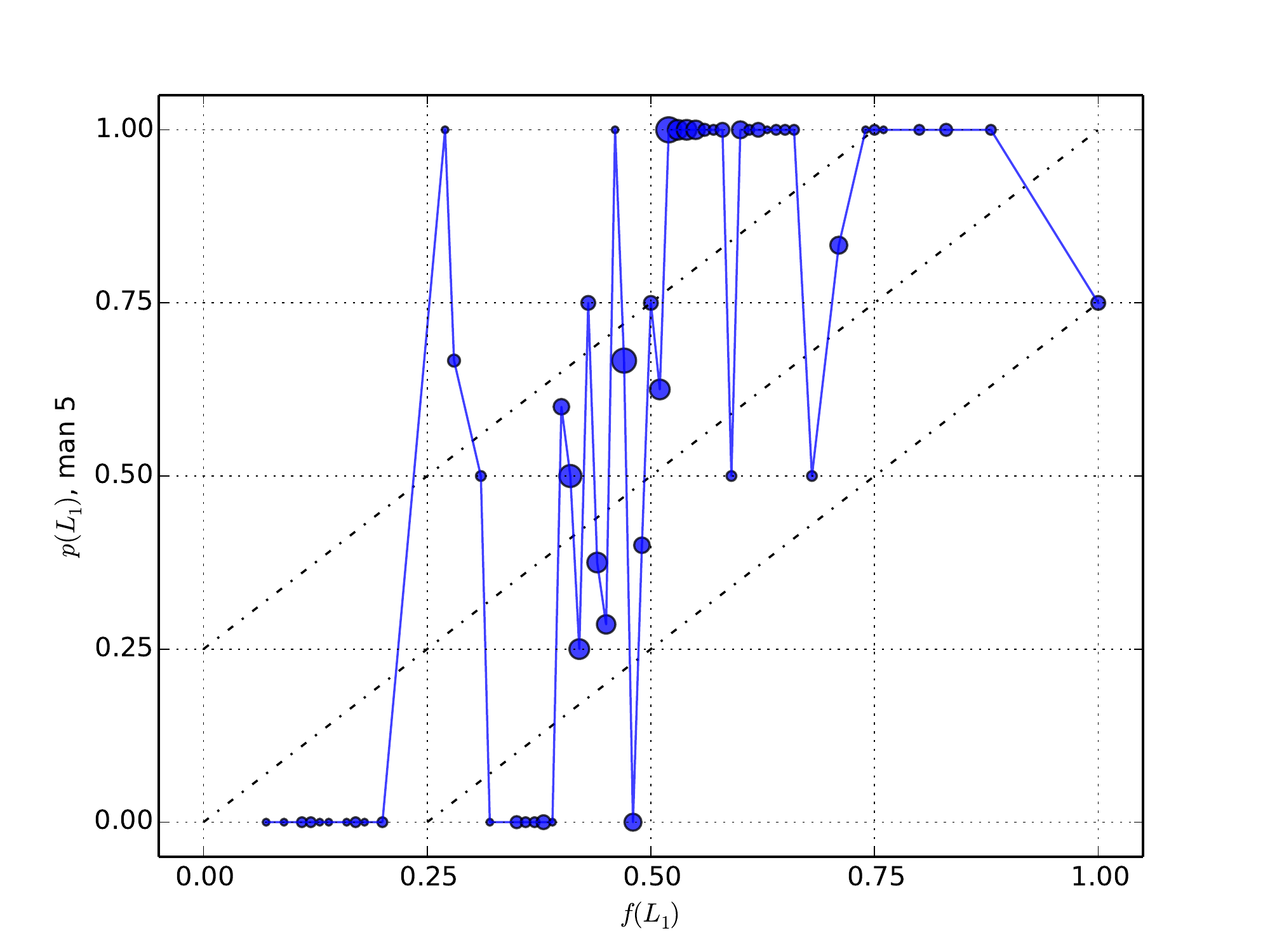}
\end{center}
\caption{For four participants separately, this figure shows empirical probability $p(L_1)$ (choice frequency of the risky prospect) as a function of the utility factor $f(L_1)$. Each point aggregates decision tasks presenting the same value of $f(L_1)$ rounded to the lowest 0.01. In each panel, the area of the markers is proportional to the number of decision tasks per point, which ranges from 1 to 12 or 13 in these cases. The diagonal lines relate to the quarter law (see equation (\ref{quarter})).}
\label{F-individuals}
\end{figure}

It is instructive to take a closer look at individual results. Fig~\ref{F-individuals} shows the results of four participants manifesting tendencies shared by other subjects in the sample. Each panel corresponds to one participant. We may make at least two general observations based on this figure. First, it is apparent that the quarter law would offer an unsatisfactory fit for these individuals, except perhaps for the one in the top right panel. To see this, one may compare Fig~\ref{QDT_prediction} to these panels. Going further, the next section will examine whether aggregating individual results helps obtain agreement with the quarter law.

Second, it appears that individuals behave quite differently from one another. For example, the participant in the top left panel appears to have calculated the expected values of each prospect and consistently chosen the one with the highest expected value. By contrast, in the top right panel, the transition from the certain to the risky prospect is noisier and takes place along a longer interval of $f(L_1)$. The participant from the bottom left panel shows a strong preference for the certain prospect until high values of $f(L_1)$. In the bottom right panel, a relatively large number of choices seem to be random, although there is overall a transition from the certain to the risky prospect. Fig~\ref{F-individuals} indicates the gender of participants, but let us stress that for each panel shown here, one may find a subject of the other gender whose choices follow a similar pattern. 

%\begin{figure}[!h]
%\caption{{\bf Individual results for four participants.}
%For four participants separately, this figure shows the empirical probability $p(L_1)$ (choice frequency of the risky prospect) as a function of the utility factor $f(L_1)$. Each point aggregates decision tasks presenting the same value of $f(L_1)$ rounded to the lowest $0.01$. In each panel, the area of the markers is proportional to the number of decision tasks per point, which ranges from $1$ to $12$ or $13$ in these cases. The diagonal lines relate to the quarter law (see Eq~(\ref{quarter})).}
%\label{F-individuals}
%\end{figure}

\subsection*{Aggregation by gender} \label{S-results-gender}

\subsubsection*{Female and male distributions of attraction factors}

In this section, we carry out an aggregation of individual results by gender and examine whether the quarter law holds at this level and whether gender differences can be observed in our sample. We use gender out of convenience, because it is a readily available characteristics of subjects that splits the sample in two sets of almost equal size. Moreover, it is common to observe gender differences in risk-taking behavior \citep{Eagly95,Byrnes99,EckelGrossman08,CrosonGneezy09}, as briefly reviewed in the discussion. 

\begin{figure}[ht!]
\centering
\includegraphics[width=0.7 \textwidth]{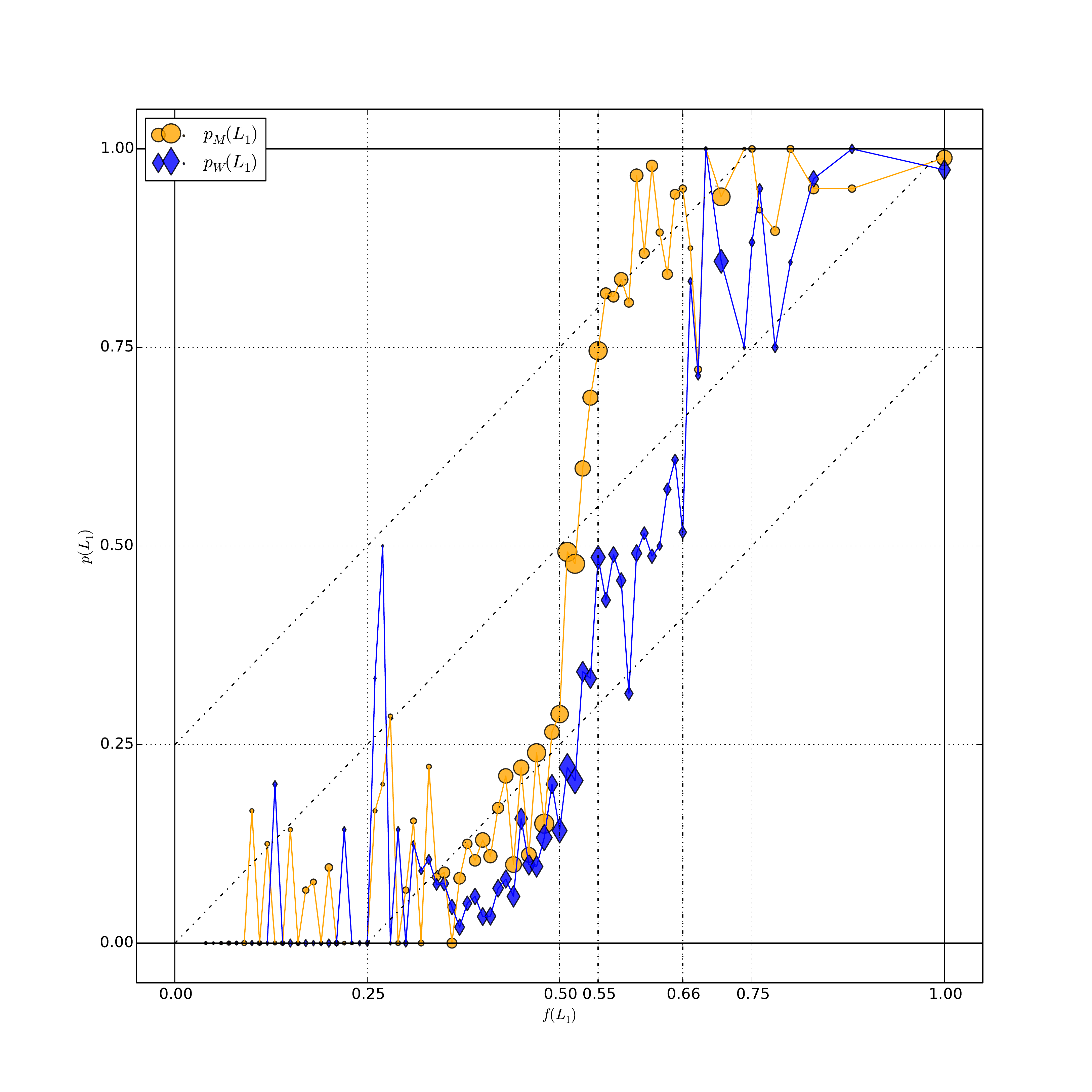}
\caption{For women (W) and men (M) separately, empirical probability $p(L_1)$ (choice frequency of the risky prospect) as a function of the utility factor $f(L_1)$. Each point aggregates decision tasks presenting the same value of $f(L_1)$ rounded to the lowest 0.01. The area of the markers is proportional to the number of decision tasks per point, which ranges from 1 to 140 for women and from 1 to 134 for men. The vertical dotted lines at $f(L_1)=0.55$ and 0.66 relate to the transition, and the diagonal lines to the quarter law (see equation (\ref{quarter})).}
\label{Gender-p}
\end{figure}

Fig~\ref{Gender-p} shows the frequency with which the risky prospect was chosen by women ($p_{W}(L_{1})$) and men ($p_{M}(L_{1})$) separately, as functions of the utility factor $f(L_{1})$.

%\begin{figure}[!h]
%\caption{{\bf Results by gender.}
%For women (W) and men (M) separately, empirical probability $p(L_1)$ (choice frequency of the risky prospect) as a function of the utility factor $f(L_1)$. Each point aggregates decision tasks presenting the same value of $f(L_1)$ rounded to the lowest $0.01$. The area of the markers is proportional to the number of decision tasks per point, which ranges from $1$ to $140$ for women and from $1$ to $134$ for men. The vertical dotted lines at $f(L_1)=0.55$ and $0.66$ relate to the transition between the two options (see text), and the diagonal lines to the quarter law (see Eq~(\ref{quarter})).}\label{Gender-p}
%\end{figure}

We perform a quantitative gender comparison using a Kolmogorov-Smirnov test on the two distributions of attraction factors $q_W(L_1)$ (for women) and $q_M(L_1)$ (for men). We use the SciPy script \texttt{\detokenize{scipy.stats.ks_2samp}} \citep{KS-scipy}. This test yields a p-value of $0.02$. As a reminder, this means that, assuming the null hypothesis that the two samples $q_W(L_1)$ and $q_M(L_1)$ are drawn from the same distribution, there is a $2$\% chance to observe the obtained empirical data. When the p-value is smaller than $0.05$, the consensus is to reject the null hypothesis and consider that the two samples, $q_W(L_1)$ and $q_M(L_1)$, are drawn from different distributions. The p-value we find thus means that the distributions of attraction factors are significantly different for men and women in our sample. 

\subsubsection*{Transition from the certain to the risky prospect} \label{S-gender-transition}

Fig~\ref{Gender-p} suggests that the preference transition from the certain to the risky prospect takes place approximately in the interval $f(L_{1}) \in [0.5,0.55]$ for men, as compared to $f(L_{1}) \in [0.55,0.66]$ for women. Namely,
\begin{itemize}
\item women (as a group) transition from the certain to the risky prospect at a higher value of $f(L_{1})$ than men, and
\item women (as a group) transition over a longer interval of $f(L_{1})$ than men.
\end{itemize}

The distance between the risk-neutral value $f(L_1)=0.5$  and the preference transition to the risky prospect reflects the population's average risk aversion. The larger distance for women suggests that women in the sample show a stronger risk aversion than men, on average. However, let us stress that the sample size of thirteen women and fourteen men is too small to generalize these results to genders in general. We come back to this point in the discussion.

The interval length, for its part, can have two different interpretations. 
\begin{itemize} 
\item If individual transitions all start and end at roughly the same values of $f(L_1)$ (homogeneous population), a longer interval reflects that decision makers show little discrimination between different values of $f(L_1)$ along that interval, and are inconsistent in their choices, leading to intermediary values of $p(L_1)$.
\item Alternatively, if individual transitions occur at different points or along intervals of different length (heterogeneous population), averaging over individuals will yield a broad transition at the level of the group. In that case, the transitional interval length at the group level reflects inter-individual variability.
\end{itemize}

The analysis performed in the next section suggests that inter-individual variability is higher among women than men and is likely to contribute to the longer transition observed in Fig~\ref{Gender-p} for women.

\subsubsection*{Intra- and inter-gender variability} \label{section-indiv-test}

We now compare the distributions of attraction factors in pairs of individuals, separating between same-gender and cross-gender pairs. We thus make three types of pairings, man-man, woman-woman and woman-man. For each pair, we perform a Kolmogorov-Smirnov test on the individual distributions of attraction factors $q(L_1)$. This yields one p-value per pair of individuals. Small p-values indicate that individuals in a pair make significantly different decisions.

\begin{figure}[ht!]
\centering
\includegraphics[width=0.7 \textwidth]{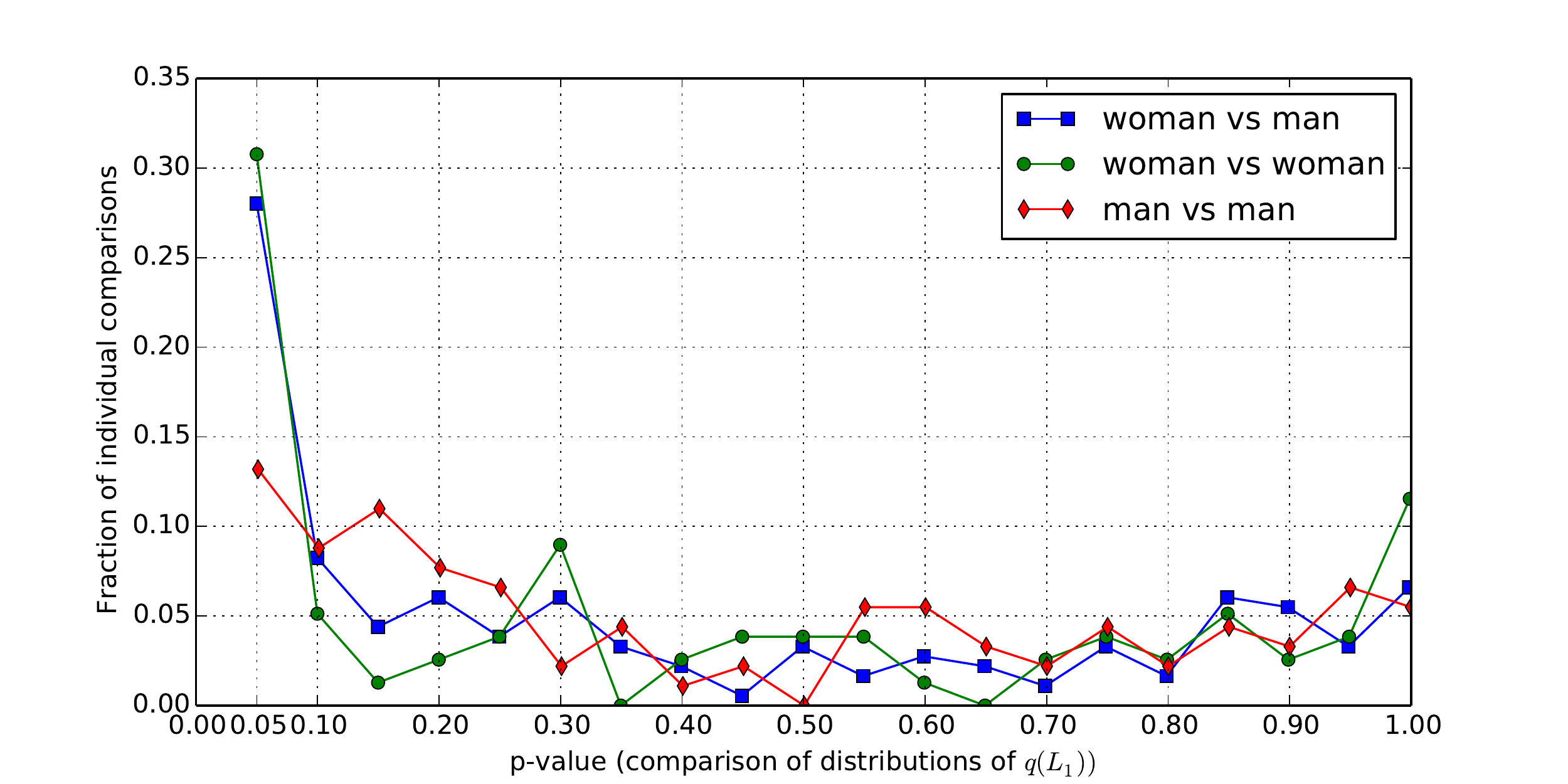}
\caption{Relative frequency of p-values (per intervals of 0.05) resulting from Kolmogorov-Smirnov tests of $q(L_1)$ distributions between pairs of decision makers. 30.1\% of the woman-woman comparisons yield a p-value between 0 and 0.05 (i.e., their $q(L_1)$ distributions are clearly distinguishable), as compared to 28\% of the woman-man comparisons and 13.1\% of the man-man comparisons.}
\label{Gender-p-values-fraction}
\end{figure}

Fig~\ref{Gender-p-values-fraction} shows the relative frequency of these p-values for the three types of pairings. About $13\%$ of the man-man p-values are below $0.05$, as compared to $28\%$ of the man-woman p-values and $31\%$ of the woman-woman p-values. 

%\begin{figure}[!h]
%\caption{{\bf Comparison of individual distributions of attraction factors.}
%Relative frequency of p-values (per intervals of $0.05$) resulting from Kolmogorov-Smirnov tests of $q(L_1)$ distributions between pairs of decision makers. $30.1$\% of the woman-woman comparisons yield a p-value between $0$ and $0.05$ (i.e., their $q(L_1)$ distributions are clearly distinguishable), as compared to $28$\% of the woman-man comparisons and $13.1$\% of the man-man comparisons.}
%\label{Gender-p-values-fraction}
%\end{figure}

We also compare these distributions of p-values, using again Kolmogorov-Smirnov tests, and find that the man-man distribution is very different from the man-woman distribution ($p=0.025$) and from the woman-woman distribution ($p=0.05$), but the woman-woman distribution is quite similar to the woman-man distribution ($p=0.68$). 

This means that the women in our sample are more dissimilar from each other than the men are; women are almost as different from each other as they are from men. In other words, the intra-group variability is greater among women than among men.

\subsubsection*{Quarter law}

In this section, we test the validity of the quarter law at the level of each gender. According to QDT, the absolute value of the attraction factor of either option in a binary lottery should be equal to $1/4$ on average. We thus compute the average of $| q(L_1) |$ over all decision tasks for women and men separately. Let us recall that one has to exclude from the analysis the decision tasks with $p(L_{1})=0$ or $p(L_{1})=1$, as explained in the methods.

%Whether the empirical average value of $| q(L_1) |$ can be said to be in agreement with the quarter law depends on whether the measure of variability covers the difference between the empirical value and the theoretical average of $0.25$. 
We may use two different aggregations of individual results to obtain the mean of $|q(L_1)|$ at the level of a group, as well as two different measures of variability. For completeness, we present all of these computations, as each gives a different perspective on the empirical results. 
 
Let us consider a sample of $N_s$ subjects to illustrate the different computations. Say that overall $N_f$ different values of $f(L_1)$ were offered to this sample. Each value of $f(L_1)$ is associated with the choices made by different individuals. These choices can be aggregated to get one value of $q(L_1)$ per value of $f(L_1)$, without differentiating between individuals. This amounts to interpreting the sample as a set of homogeneous individuals, or equivalently as a single entity. We then average these $N_f$ values of $|q(L_1)|$ ($N_f$ lies between $50$ and $60$ in the calculations presented below). The standard deviation measures the variation of $|q(L_1)|$ across all decisions and sample subjects aggregated.

%In general, for each subject, we aggregate the decisions presenting the same utility factor $f(L_1)$. Thus, if a subject $s_i$ is exposed to $n_i$ values of $f(L_1)$, then $s_i$ is associated with $n_i$ values of $|q(L_1)|$. A first approach is to put together these $\sum_i^{N_s} n_i$ values for all sample subjects and compute the standard deviation of these values. The standard deviation thus measures the variation of $|q(L_1)|$ across all decisions and sample subjects aggregated.
%We thus get several values of the absolute attraction factor $|q(L_1)|$ for each subject, that is, one value of $|q(L_1)|$ for each value of $f(L_1)$ that the subject is presented with.

Another approach is to first isolate each individual's choices so as to obtain the subject's average value of $|q(L_1)|$ over the different values of $f(L_1)$ submitted to this subject alone. We thus get $N_s$ values of $\overline{|q(L_1)|}$, one for each subject, and then average these. Their standard deviation measures the variation across subjects in the sample. This standard deviation is smaller than the previous one, because we first average at the individual level and thus remove within-individual variation. The sample mean $\overline{|q(L_1)|}$ may also be different as in the previous computation. 

An alternative measure of variability to the standard deviation is the standard error of the mean (SEM). The SEM (that can only be estimated here) measures the variation between same-size samples compared to the full population from which the sample is drawn. Here, the ``full population" is not the world human population. It is the comparable population in terms of age, gender (if the sample is restricted to one gender), socio-economic situation, education level, and culture. It may be qualified as WEIRD (``western, educated, industrialized, rich, and democratic"), as coined by \citet{Henrichetal-WEIRD}. The estimated SEM is equal to the standard deviation divided by the square root of the number of values ($N_f$ or $N_s$) whose variability is being measured and thus depends on the chosen aggregation.

These approaches yield the following attraction factors, standard deviations and SEM (in brackets), for women (W) and men (M). Without differentiating between subjects, we get 
%With $n_i$ values of $|q(L_1)|$ per subject, we get
% n_i \textrm{ values of } |q(L_1)| \textrm{ per subject: } 
% 1 \textrm{ value of } \overline{|q(L_1)|} \textrm{ per subject: } 
\begin{equation} \label{e-q-res-1}
\overline{|q_\textrm{W}(L_1)|} = 0.20 \pm 0.12 \, (0.02),
\end{equation}
\begin{equation}  \label{e-q-res-2}
\overline{|q_\textrm{M}(L_1)|} = 0.19 \pm 0.10 \, (0.01).
\end{equation}

With a single value of $\overline{|q(L_1)|}$ per subject, we get
\begin{equation}
\overline{|q_\textrm{W}(L_1)|} = 0.21 \pm 0.05 \, (0.01),
\end{equation}
\begin{equation} \label{e-q-res-4}
\overline{|q_\textrm{M}(L_1)|} = 0.20 \pm 0.04 \, (0.01).
\end{equation}

%mean, std, standard error 1/sqrt(N) * std
%ddof_param =  0 (std with 1/N if ddof=0, 1/(N-1) if ddof=1) 
%FEMALE INDIV first, mean(abs(q1)) =  0.21  +/-  0.05 0.01
%MALE INDIV first, mean(abs(q1)) =  0.2  +/-  0.04 0.01
%INDIV first, mean(abs(q1)) =  0.21  +/-  0.05 0.01
%* mean(abs(q1)) =  0.17  +/-  0.1 0.01 ( 60 )
%* mean(abs(q1_wom)) =  0.2  +/-  0.12 0.02 ( 51 )
%* mean(abs(q1_men)) =  0.19  +/-  0.1 0.01 ( 52 )

In all cases, the mean value of $|q(L_1)|$ is reasonably close to the value of $0.25$ predicted by the quarter law, and this is formally evaluated by the standard deviation or SEM. When we consider each aggregate as an entity and use the standard deviation, the results for both genders are in agreement with the quarter law prediction of $0.25$ on average. We interpret in the discussion the relatively large standard deviations found in this case. 

When we average over each subject first and thus get a single value of $\overline{|q(L_1)|}$ per subject and consider the standard deviation, the results are also in agreement with the quarter law at the $95\%$ confidence level. This corresponds to a difference between the empirical and the theoretical value lower than approximately two (not one) standard deviations. This is because $95\%$ of a Gaussian distribution lies within $1.96$ standard deviations on both sides of the mean.

%as long as we use the one standard deviation difference criterion. Indeed, one usually rejects a model at the $95\%$ confidence level only when the difference between the empirical and the theoretical value is more than two (not one) standard deviations. This is because $95\%$ of a Gaussian distribution lies within roughly $1.96$ standard deviations of the mean. 

With both approaches, the SEM does not cover the difference between the empirical and the theoretical value. This can be expected because, as noted above, the sampled population is biased and thus does not satisfy the no prior information hypothesis at the basis of the quarter law. We briefly come back to this point in the discussion.

Overall, these results indicate that individual choices vary sufficiently to span the difference between the empirical value of $|q(L_1)|$ and the theoretical value of $0.25$. However, the sample is overall biased. It could be the case that a larger sample would bring the mean absolute of the attraction factor nearer to $0.25$ so that the SEM would cover the difference. This would occur if the bias of the mean decreases faster than the SEM as the sample size increases. Yet this is rather unlikely, as more of the same ``kind" of subjects (in terms of socio-economic background, etc.) would still constitute a biased sample. 

\subsection*{General population: quarter law}

In this section, we aggregate individual results with the aim to test the quarter law at the population level.

First, we check the prediction given by Eq~(\ref{e-sign-q}) applying to the situation where the utility factors of the two competing prospects are very close. At $f(L_1)=0.5$, we find the following attraction factor, based on $224$ decision tasks:
\begin{equation}
\bar{q}_{\textrm{at}}(L_1) = -0.29~.
\label{jujuywg}
\end{equation}
Within an experimental error of $14\%$, this result is in agreement with the quarter law and confirms the rule specifying the sign of the attraction factor close to $f(L_1) = 0.5$, expressed by \citet{QDTManip14} (p.1162) and recalled in Eq~(\ref{e-sign-q}).

\begin{figure}[ht!]
\begin{center}
 \includegraphics[width=.7 \textwidth]{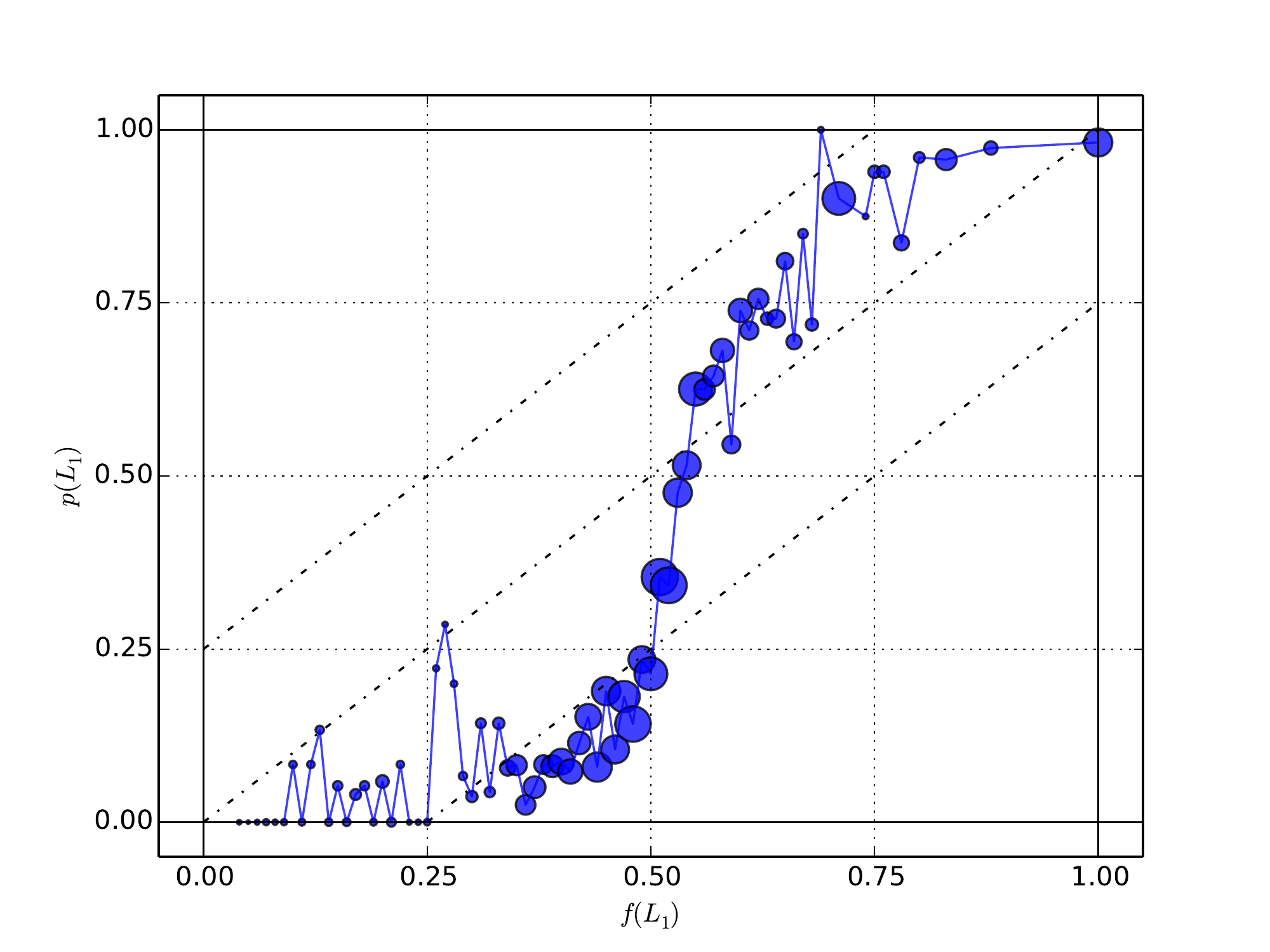}
\end{center}
\caption{Empirical probability $p(L_1)$ (choice frequency of the risky prospect) as a function of the utility factor $f(L_1)$. Each point aggregates decision tasks presenting the same value of $f(L_1)$ rounded to the lowest 0.01. The area of the markers is proportional to the number of decision tasks per point and varies between 3 and 274.} 
\label{Same-f-p1-f1}
\end{figure}

Fig~\ref{Same-f-p1-f1} shows $p(L_1)$ as a function of $f(L_1)$ for the complete population. 

%\begin{figure}[!h]
%\caption{{\bf Population results.}
%Empirical probability $p(L_1)$ (choice frequency of the risky prospect) as a function of the utility factor $f(L_1)$ for the complete population. Each point aggregates decision tasks presenting the same value of $f(L_1)$ rounded to the lowest $0.01$. The area of the markers is proportional to the number of decision tasks per point and varies between $3$ and $274$.}
%\label{Same-f-p1-f1}
%\end{figure}

As explained in the methods, one has to exclude from the analysis the decision tasks with $p(L_{1})=0$ or $p(L_{1})=1$. We repeat the procedure used when we tested the quarter law for each gender (Eqs~(\ref{e-q-res-1})-(\ref{e-q-res-4})). We thus compute the mean of $|q(L_1)|$ and the standard deviation in the different ways described earlier and also estimate the population standard error of the mean (in brackets). Without differentiating between subjects, we get
\begin{equation} \label{e-pop-q}
\overline{|q(L_1)|} = 0.17 \pm 0.10 \, (0.01), 
\end{equation}
and with a single value of $\overline{|q(L_1)|}$ per subject, we get
\begin{equation}
\overline{|q(L_1)|} = 0.21 \pm 0.05 \, (0.01).
\end{equation}

%Note that the result obtained on the basis of $n_i$ values per subject is lower than the two results obtained separately for women and men on the same basis (Eqs~(\ref{e-q-res-1})-(\ref{e-q-res-2})). This is because here female and male decisions with the same rounded value of $f(L_1)$ are aggregated 

Considering the standard deviation, these results are in agreement with the quarter law prediction of $0.25$ on average. We interpret in the discussion the relatively large standard deviation found in Eq~(\ref{e-pop-q}). As in the gender analysis, the SEM does not cover the difference between empirical and theoretical value, and this is likely due to the population being biased. 

%mean, std, standard error 1/sqrt(N) * std
%ddof_param =  0 (std with 1/N if ddof=0, 1/(N-1) if ddof=1) 
%FEMALE INDIV first, mean(abs(q1)) =  0.21  +/-  0.05 0.01
%MALE INDIV first, mean(abs(q1)) =  0.2  +/-  0.04 0.01
%INDIV first, mean(abs(q1)) =  0.21  +/-  0.05 0.01
%* mean(abs(q1)) =  0.17  +/-  0.1 0.01 ( 60 )
%* mean(abs(q1_wom)) =  0.2  +/-  0.12 0.02 ( 51 )
%* mean(abs(q1_men)) =  0.19  +/-  0.1 0.01 ( 52 )

\section*{Discussion} \label{S-discussion}

\subsection*{Characterization of individual decision makers} \label{S-discuss-indiv} 

Our analysis suggests that the quarter law holds at the group level and could be refined at the individual level to reflect characteristics of decision makers. As noted during the derivation of the quarter law leading to Eq~(\ref{quarter}), the value $1/4$ is based on the condition that there is no prior or additional information other than the presentation of the two competing prospects. Physiobiological influences, psychological or cultural framing can be expected to break this hypothesis and give different predictions for the attraction factor, as shown in specific cases by \citet{QDTManip14,QDTInfoSocial14}.

One may thus ask how the attraction factor can be reliably used to characterize individuals. That is, one may attempt to calibrate the attraction factor on an empirical dataset for a given individual and make predictions for the same individual's decisions at a later time in a similar context. This is currently examined in prospect theory \citep{GonzalezWu99, FehrDuda12, Murphy14}. Since QDT is a probabilistic theory, one would get a prediction for the frequency with which an individual chooses an prospect. 

One shall thus examine specific forms for the distribution of attraction factors $q(L_j)$, and possibly refine the form of the utility factor $f(L_j)$ by replacing the expected values $U(L_j)$ (Eqs~(\ref{utility1})-(\ref{utility2})) by expected utilities, as for example in prospect theory \citep{1979KT,1992TverskyKahn} (see Eq~(\ref{utility_PT})). This may lead to an approximation for the form of $p(L_j)$ approaching a logistic function, as suggested by Fig~\ref{Same-f-p1-f1}, which is done in classical approaches (e.g. \citealp{MostellerNogee51, Murphy14}). Such fitting is beyond the scope of the present work, whose main objective is to provide a guide of how to apply QDT in its present theoretical state to a dataset of simple gamble questions, but is the focus of current and future research \citep{Calibration2016}.

\subsection*{Gender differences and aggregation of individual data} \label{S-discuss-gender}

There is a vast documentation on gender differences: reviews include \citet{Byrnes99,EckelGrossman08} and \citet{CrosonGneezy09}. Financial risk preferences in particular are studied by \citet{Powell97,Schubert99} and \citet{FehrDuda06}, among others. Understandably, this is a socially and politically loaded topic; see in particular the survey by \citet{Eagly95}. In fact, it is next to impossible to find any research article that does not insist on being cautious in its conclusions (and we are no exception, as it happens). Reasons given are often the small sample size, possible biases from various demographic variables not controlled for, or artificial settings. 

Despite some notable exceptions \citep{Daruvala07}, the current state of affairs points toward the idea that there are indeed gender differences in risk-taking behavior, namely that women exhibit on average more risk aversion than men, with overlapping distributions. It has been suggested that gender differences in risk-taking behavior depend heavily on demographic factors: \citet{Hibbert08} surveyed 1,382 finance and English professors and found that, when individuals have the same level of education, women are no more risk averse than their fellow male colleagues. \citet{FehrDuda06} suggest that women and men have different probability weighting schemes that lead to a more risk averse behavior on the part of women. Gender differences in risk-taking behavior has been linked to the sociobiologists' hypothesis that men take more risks when they are trying to attract mates, while in parallel women tend to be more risk averse during their childbearing years (see \citealp{Brinig95} p.17 n61, and \citealp{FavreSornette12}).

In our dataset, men appear to show on average less risk aversion than women. Another observation is that women show a greater inter-individual variability than men. Yet, the gender comparison we perform comes with a disclaimer about the small sample size. Thirteen women and fourteen men cannot be taken as representative of the genders. Our purpose, rather than to draw conclusions about gender differences in risk preferences, has been to illustrate how to apply and test QDT at different levels, namely individuals and groups. Gender is a variable that is usually readily available and there is a vast literature on gender-specific risk-taking behavior, as briefly reviewed above. This makes gender an obvious first candidate on which to test the aggregation of individual data. 

It is clear from Fig~\ref{F-individuals} that not only individuals differ significantly from each other in their decision-making strategies, but additionally within-individual variability can be quite high. The aggregation of decisions made by different individuals thus results in high standard deviations, as seen in this study. Although this is not always stressed in analyses of decision making experiments, high levels of stochasticity appear to be a feature of human decision making in general and not of the dataset analyzed here in particular. 

When looking at individual panels such as those of Fig~\ref{F-individuals}, it seems possible that individuals may follow a relatively small number of distinct strategies. In other words, individuals may manifest a limited number of thinking or decision-making schemes (see e.g. \citealp{Poncela-Casasnovas2016, Slovic16, DouglasWildavskyRisk, FavreThesis}). For example, some participants may make random choices, under the belief that the lottery is rigged or that they are generally unlucky. Other participants may take calculated risks according to set rules, possibly leading them to systematically select the option with the highest expected value. Some may shun the risky option in almost all cases, independently of the amounts and probabilities involved. A number of people may also avoid these rather extreme possibilities and more often than not choose the option with the highest expected value while still exhibiting a certain degree of stochasticity.

If such a clustering of strategies could be identified, it would potentially provide a meaningful and reliable criterion for the characterization and aggregation of individual data and help set the limits of the predictability of individuals. That is, individuals could be partly characterized in terms of being more or less predictable (at least in a certain context and at a certain time), and the inability of a model to predict individual decisions precisely may not be interpreted as a failure of the model. The quarter law, based on the no prior information hypothesis, could be expected to hold in terms of the standard error of the mean when a sample includes subjects showing all types of strategies. Again, this article offers a practical guide on how to analyze binary lottery data in the domain of gains using QDT, but does not address the problem of identifying such generic strategies. This is the focus of future research and is beyond the scope of the present article. 

\subsection*{Lotteries generation and sample size} \label{S-design}

One of the practical challenges we encountered was to choose an adequate aggregation of decision tasks in order to carry out a meaningful probabilistic analysis guided by QDT, given the empirical distribution of $(p,x)$ decision tasks presented to subjects. In future experiments meant to be analyzed with QDT, we suggest the following lottery generation method. Let $N$ be the number of decision tasks presented to each participant. Each decision task presenting a given value of the utility factor $f(L_j)$ should be planned to be offered at least three times to the same participant, and preferably more in the transition interval $f(L_j) \in [0.4,0.7]$. A rounding of $f(L_j)$ should be chosen such as to allow this repetition over the N decision tasks and have the distribution of $f(L_j)$ cover the interval $[0,1]$. The lotteries should be generated beforehand and the distribution of games should be controlled to reflect the desired distribution. The order in which $(p,x)$ games are presented to subjects should be randomized so that participants are unlikely to answer very similar questions consecutively.

Finally, $27$ subjects do not constitute a large enough sample to generalize our results to a broader population, despite the large number of choice problems per subject ($200$). As a comparison, prospect theory founders Kahneman and Tversky typically ran experiments on $50$ to $300$ subjects \citep{1979KT, TK1983, 1992ShafirTversky}. The methodology presented in this work should be replicated in experiments with more participants (and more diverse ones; \citealp{Henrichetal-WEIRD}) in order to refine the QDT quantitative analysis of the observed phenomena.

\section*{Conclusion} \label{S-conclusion}

We offer in this work a guide and illustration of how to apply quantum decision theory (QDT) to an empirical dataset of binary lotteries in which decision makers had to choose between a certain and a risky option involving gains. Through this, we hope to promote the empirical investigation of QDT, in order to test its current claims, but also challenge it and indicate new research directions. One such direction could be to calibrate the attraction factor and characterize individual decision makers in terms of type of strategy and degree of predictability. 

\section*{Supporting Information}

% Include only the SI item label in the paragraph heading. Use the \nameref{label} command to cite SI items in the text.
\paragraph*{S1 File.}
\label{S1_File}
{\bf Dataset.} Each row in the dataset represents a decision task and includes an individual identifier ($1$ to $27$), the gender of the participant (F or M), the probability with which the amount CHF $50$ would be obtained in the risky option, the amount guaranteed by the certain option, and the decision made by the participant ($0$ for the certain option, and $1$ for the risky option). There are $200$ decision tasks per participant, resulting in $5,400$ data rows.

\section*{Acknowledgments}

For constructive discussions and feedbacks, we thank Alexandru Babeanu, Beno\^it Crouzy, George Harras, Tatyana Kovalenko and Christine Sadeghi. We also thank anonymous reviewers for their careful reading and helpful comments that contributed to improve this article. %We acknowledge the Swiss National Foundation (grant 2-77095-15) for partial funding. %todo this goes elsewhere,

% Either type in your references using
% \begin{thebibliography}{}
% \bibitem{}
% Text
% \end{thebibliography}
%
% or
%
% Compile your BiBTeX database using our plos2015.bst
% style file and paste the contents of your .bbl file
% here.
% 

\bibliographystyle{ametsoc} 
\bibliography{RALT_Data_Analysis}

\end{document}